\documentclass{aa}

\usepackage{graphicx}
\usepackage{txfonts}

\usepackage[dvipsnames]{xcolor}
\usepackage{xspace}
\usepackage{anyfontsize}
\usepackage{comment}
\usepackage{amsmath}
\usepackage[]{hyperref}

\newcommand{\decode}{\textsc{decode}\xspace}
\defcitealias{fu_2025}{F25}

\begin{document}

   \title{The role of black hole feedback on galaxy star formation and the degeneracy with halo quenching}
   \titlerunning{Galaxy-black hole growth}
   \authorrunning{H. Fu et al.}


    \author{Hao Fu\inst{1,2},
            Francesco Shankar\inst{2},
            Feng Yuan\inst{1},
            Daniel Roberts\inst{2},
            Lumen Boco\inst{3},
            Andrea Lapi\inst{4},
            Pablo Corcho-Caballero\inst{5},
            Mohammadreza Ayromlou\inst{6},
            Antonis Georgakakis\inst{7},
            Brivael Laloux\inst{8},
            Iv\'an Mu\~noz Rodr\'iguez\inst{9},
            Yingjie Peng\inst{10,11}
          }

   \institute{Center for Astronomy and Astrophysics and Department of Physics, Fudan University, Shanghai 200438, China\\
              \email{haofu@fudan.edu.cn}\and
            School of Physics and Astronomy, University of Southampton, Highfield, Southampton, SO17 1BJ, UK\and
            Universit{\"a}t Heidelberg, Zentrum f{\"u}r Astronomie, Institut f{\"u}r theoretische Astrophysik, Albert-Ueberle-Str. 2, 69120 Heidelberg, Germany\and
            SISSA, Via Bonomea 265, 34136 Trieste, Italy\and
            Kapteyn Astronomical Institute, University of Groningen, PO Box 800, 9700 AV Groningen, The Netherlands\and
            Argelander-Institut f\"ur Astronomie, Auf dem H\"ugel 71, D-53121 Bonn, Germany\and
            Institute for Astronomy and Astrophysics, National Observatory of Athens, V. Paulou \& I. Metaxa, Pendeli, 11532, Greece\and
            INAF - Osservatorio Astronomico di Capodimonte Salita Moiariello 16, 80131, Napoli, Italy\and
            IRFU, CEA, Université Paris-Saclay, 91191 Gif-sur-Yvette, France\and
            Department of Astronomy, School of Physics, Peking University, 5 Yiheyuan Road, Beijing 100871, China\and
            Kavli Institute for Astronomy and Astrophysics, Peking University, 5 Yiheyuan Road, Beijing 100871, China
             }

   \date{Received September 15, 1996; accepted March 16, 1997}


%

  \abstract
   {}
   {The interplay between the accretion of supermassive black holes (SMBHs) and the stellar mass growth of the host galaxies is still a matter of hot debate. The accretion of the central SMBHs is expected to release energy under the form of active galactic nuclei. This energy is believed to impact the star formation activity and contribute to the quenching of the host galaxies. Here, we address this key unsolved issue with our cosmological semi-empirical model \decode (Discrete statistical sEmi-empiriCal mODEl).}
   {In \decode, we grow galaxies with their star formation rate linked to halo accretion rate distributions via abundance matching. SMBHs are evolved following the stellar mass growth of their host galaxies by assigning an accretion rate at each redshift from the empirical Eddington ratio distributions and duty cycles. We tested the assumption that galaxies permanently quench when their central SMBHs approach the limit imposed by the observed $M_{\rm BH} - \sigma_\star$ relation, as a proxy of SMBH disruptive feedback.}
   {We find that simply imposing the $M_{\rm BH} - \sigma_\star$ condition is sufficient to generate a fraction of quenched galaxies consistent with current data, including the newest ones from Euclid. In addition, our minimal data-driven model also predicts SMBH scaling relations consistent in slope and normalisation with those that have been observed, and an $M_{\rm BH} - M_\star$ relation weakly evolving with redshift. The model also naturally generates SMBH accretion rates peaking within 1 Gyr of their host star formation histories. Interestingly, we note that all the main predictions on galaxy quenched fractions and SMBH growth histories and scaling relations are degenerate with those expected in a halo quenching model.}
   {The comprehensive data-driven model presented in this work represents an invaluable tool to investigate SMBH demography across time and environments in an accurate, physically motivated manner, ideally suited to rapidly exploring the implications from large surveys, such as Euclid and Rubin-LSST.}


   \maketitle
%

\section{Introduction}\label{sec:intro}

There is ample evidence supporting the existence of supermassive black holes (SMBHs) living at the centre of almost all local galaxies. Several observations have reported the existence of scaling relations between SMBH masses and the properties of their host galaxies. For example, power law correlations between the SMBH mass and the galaxy bulge stellar mass and stellar velocity dispersion have been proposed by many groups (e.g. \citealt{kormendy_richstone_1995, magorrian_1998, ho_1999, haring_2004, ferrarese_2005, graham_2007, gultekin_2009, sani_2011, kormendy_ho_2013, shankar_2016, shankar_2025}). Studying scaling relations plays a pivotal role in unravelling the dynamic relationship between the growth of galaxy properties and the mass of SMBHs. Such scaling relations offer valuable insights into a range of research areas in galaxy-SMBH co-evolution, including the regulation of materials (such as gas and dust) within galaxies through active galactic nucleus (AGN) feedback, the interplay between black hole growth and the galaxy star formation rate (SFR), which in turn is linked to the availability of cold gas, and even the morphology of galaxies (see, e.g. \citealt{marconi_2008, volonteri_2013, heckman_2014, calvi_2018, king_2019}). Furthermore, the correlations between SMBHs and galaxies established in the local Universe provide a reference point for investigating the evolution of these relationships at higher redshifts (e.g. \citealt{bennert_2011, sexton_2019, pacucci_2023, li_2025}). SMBHs are believed to co-evolve with their galactic hosts and, whilst accreting gas, to shine as AGN emitting strong winds, jets, and radiative outputs at all wavelengths. The strong feedback from AGN may be sufficiently strong to reheat and expel surrounding gas up to kpc scales, though a clear consensus on these vital matters is not yet available (e.g. \citealt{croton_2006, lapi_2006, silk_2012, lapi_2014, vogelsberger_2014, schaye_2015, choi_2017, lapi_2018, weinberger_2018, yuan_2018, zinger_2020, ma_2022, piotrowska_2022, ayromlou_2023, bluck_2023}).

From the theoretical side, black hole-galaxy co-evolution has been widely studied in hydrodynamical simulations (e.g. \citealt{schaye_2015, pillepich_2018_MstarContent, yuan_2018, dave_2019}) and semi-analytical models (e.g. \citealt{lacey_2016, fontanot_2020, ayromlou_2021, spinoso_2023, cattaneo_2025}). Hydrodynamical simulations are an extremely promising tool to understand the underlying physics involved in galactic evolution thanks to their first-principle modelling, but these simulations have high requirements in terms of computational resources. Semi-analytical models are more flexible tools thanks to their analytical nature without the need of resolving full non-linear equations of the involved physics, although still characterised by a higher number of assumptions and associated input parameters.

Semi-empirical models have been conceived as a complementary tool to bridge the gap between observations and theoretical expectations (e.g. \citealt{hopkins_2009, rodriguez_puebla_2017, moster_2018, behroozi_2019, grylls_2019, zhang_2023, fu_2022, fu_2025, boco_2023, lapi_2025}). Semi-empirical models marginalise over the complex physics of galaxy formation by exploiting data-driven scaling relations between some galactic and halo properties to grow galaxies and their central SMBHs following their progenitor dark matter haloes. Semi-empirical models benefit from the nearly absence of mass resolution and a low number of initial parameters. In recent years, a number of groups (e.g. \citealt{yang_2018, georgakakis_2019, shankar_2020, munoz_rodriguez_2023, terrazas_2024, zou_2024, guetzoyan_2025}) have started developing semi-empirical models that also trace the evolution of SMBHs. In more refined instances, SMBHs are allowed to grow along their host galaxy evolutionary tracks via empirical mass accretion rate distribution functions that assign at each stellar mass and redshift a SMBH accretion rate that, integrated in time, yields the black hole mass. This approach by design ensures that 1) at each epoch the model is consistent with the observed AGN luminosity function, given by the convolution of the accretion rate and stellar mass functions, and 2) the growth of SMBHs fully reflects current observations. Alternative recent approaches to model black hole growth from a semi-empirical perspective are based on simultaneously parametrising the galaxy SFR, black hole accretion rate (BHAR) distributions and galaxy-black hole scaling relations (e.g. \citealt{zhang_2023}).

Here, we adopt a similar approach to that of \citet{georgakakis_2019}, to grow SMBHs using their observed accretion rate distributions but on top of galaxies formed via empirical SFR distributions. In the previous paper of this series (\citealt[][F25 henceforth]{fu_2025}), we preferentially focused on halo quenching from a full data-driven approach taking advantage of our state-of-the-art semi-empirical model \decode, the Discrete statistical sEmi-empiriCal mODEl, built around a monotonic connection between SFR and halo accretion rate (HAR; Sect. \ref{sec:decode_AM}), as also seen in the latest hydrodynamic simulations. In this work, we use \decode to empirically seed and grow SMBHs by linking them to the stellar mass assembly of their host galaxy via observed accretion rate distributions. Our novel data-driven model allows to grow galaxies and SMBHs in lock-step along the dark matter assembly histories guided only by observational results, namely the SFR functions and the Eddington ratio distributions. Our method also allows to implement and transparently test, in a minimal approach, different forms of quenching applied to star-forming galaxies. In \citetalias{fu_2025}, we studied the impact of simple halo quenching, where galaxies halt their star formation when overcoming a given host halo mass threshold. In this work we instead investigate the impact of SMBH quenching by allowing star-forming galaxies to halt their star formation when their central SMBHs surpass a certain mass threshold set by a predefined SMBH-galaxy scaling relation, as inspired by theoretical and observational results. The approach put forward in this work has the favourable features of 1) being fully data-driven in the evolution of its baryonic components and 2) relying on the least possible number of theoretical assumptions and parameters.

This paper is structured as follows. In Sect. \ref{sec:decode} we describe our methodology, the abundance matching and the way we grow our SMBHs. In Sect. \ref{sec:results} we show our results on the black hole accretion tracks and galaxy-black hole coevolution. Finally, in Sects. \ref{sec:discuss} and \ref{sec:conclu} we discuss our results and draw our conclusions. Throughout we adopt the $\Lambda$CDM cosmology\footnote{We corrected all datasets to the same cosmology used in this work. We note, however, that the small differences in the cosmological parameters have a negligible impact on our results.} with best fit parameters from \citet{planck2018_cosmo_params} (i.e. $(\Omega_{\rm m}, \Omega_\Lambda, \Omega_{\rm b}, h, n_{\rm S}, \sigma_8) = (0.31, 0.69, 0.049, 0.68, 0.97, 0.81)$), and all the input and reference datasets adopted here are based on a \citet{chabrier_2003} stellar initial mass function.

\section{The DECODE implementation}\label{sec:decode}

\begin{figure*}
    \centering
    \includegraphics[width=\textwidth]{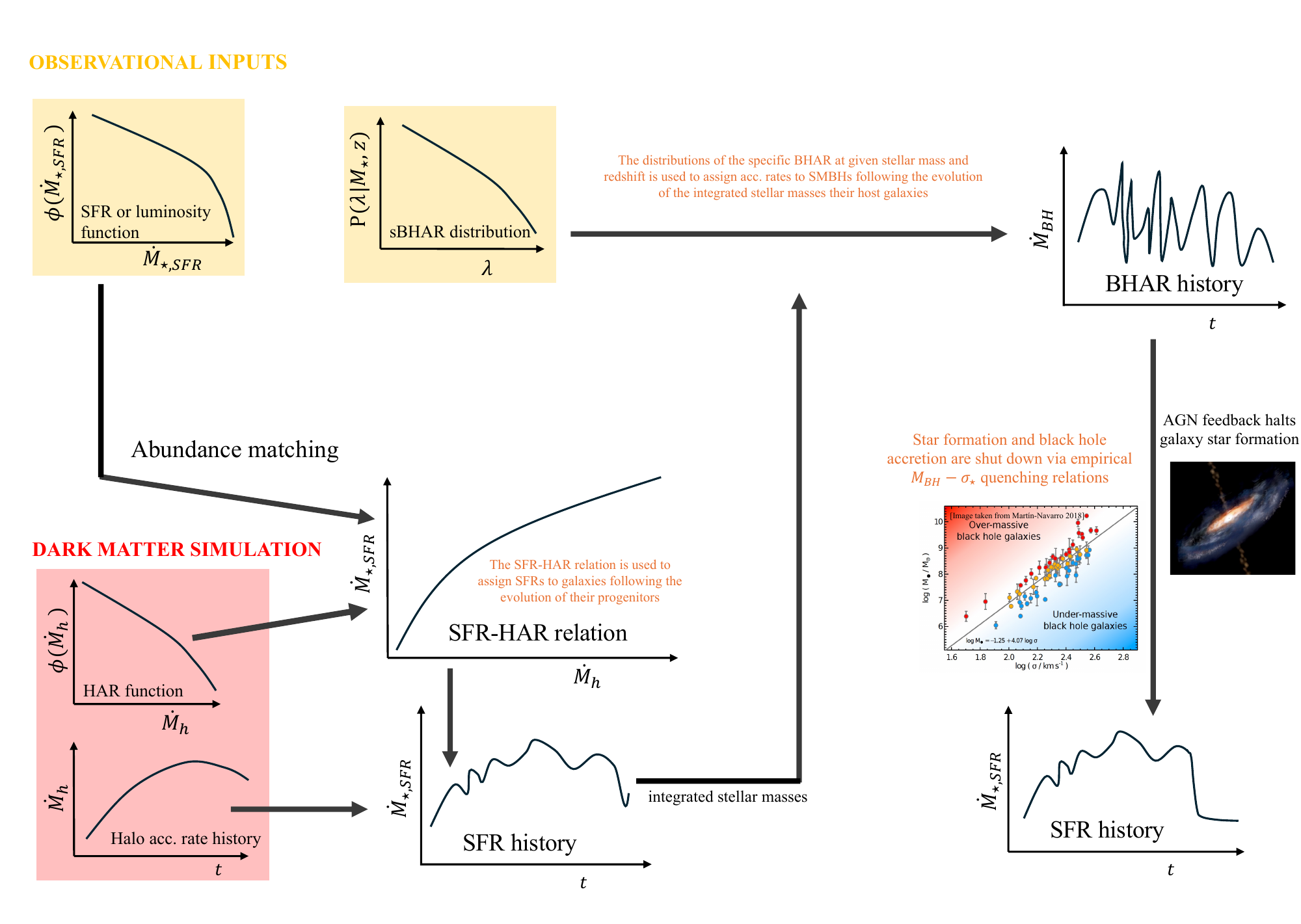}
    \caption{Methodology followed in \decode to form and evolve galaxies and supermassive black holes. The SFR-HAR relation, computed via abundance matching, is used to assign galaxies following the accretion histories of their host dark matter haloes. Observationally determined sBHAR distributions at each redshift are used to assign the accretion rates to the central SMBHs following the stellar mass growths of their host galaxies. Black hole accretion and star formation rate are halted using empirically inferred $M_{\rm BH} - \sigma_\star$ quenching relations.}
    \label{fg:cartoon}
\end{figure*}

In this section we describe our step-by-step semi-empirical and data-driven approach to grow galaxies and their central SMBHs along their host halo assembly histories. Our method, incorporated into the semi-empirical model DECODE\footnote{\href{https://github.com/haofuastro/DECODE2}{https://github.com/haofuastro/DECODE2}}, starts by assigning SFRs to galaxies from a monotonic relation between SFR and HAR derived from abundance matching (see below). The latter relation is fully supported by the predictions of cosmological simulations, such as TNG and EAGLE (\citealt{schaye_2015, pillepich_2018_MstarContent}), as proven in \citetalias{fu_2025}. This approach allows one to self-consistently grow star-forming galaxies, but additional assumptions are needed to build-up the population of quenched galaxies. In \citetalias{fu_2025} we explored the simple case of halo quenching and found that this condition is sufficient to provide a reasonable match to the quenched fractions at redshift $z<3$. Here we go a step further and explore the impact of SMBH quenching. To this purpose, we first self-consistently grew SMBHs in our semi-empirical framework, and then assumed that galaxies quench when the SMBH reaches a certain mass threshold, set by a given scaling relation.

The main steps of our methodology are summarised as follows:
\begin{itemize}
    \item computing the SFR-HAR relation;
    \item assigning SFRs to galaxies;
    \item computing the integrated stellar mass growth;
    \item assigning accretion rates to the central SMBHs from measured stellar mass- and redshift-dependent Eddington ratio distributions;
    \item integrate accretion rate distribution along the progenitor histories to retrieve the SMBH masses at each cosmic time;
    \item quenching star formation in the host galaxies when the central SMBH passes a pre-defined mass threshold.
\end{itemize}

Figure \ref{fg:cartoon} presents an illustration of our methodology. As described in \citetalias{fu_2025}, we started from the SFR-HAR relation computed via abundance matching between the SFR and HAR functions. We then used the SFR-HAR relation to assign galaxies a SFR at any given epoch based on the HAR of their host halo. By following the HAR evolutionary history, we could then build the star formation history of its central galaxy. We then grew central black holes starting from a given initial condition at high redshift (see Sect. \ref{sec:grow_SMBHs}) and assigned to each black hole a probability to be active and an accretion rate at each redshift from the input accretion rate distributions. We then integrated in cosmic time along the progenitor evolutionary tracks both the SFR and SMBH accretion rates to retrieve the galaxy and SMBH masses, which in turn yielded information on the demography and scaling relations of SMBHs and galaxies. We detail each step of our method in the following.

    \subsection{Dark matter halo catalogues}\label{sec:decode_DM_halo_cat}

    \decode is a semi-empirical model that forms and evolves galaxies on top of dark matter assembly histories, taking in input both their halo accretion and merger rates. As carried out in \citetalias{fu_2025}, we base our analysis on a catalogue of $N=10^5$ central dark matter haloes with mass above $M_{\rm h} > 10^{11} \, M_\odot$ at each redshift of interest. These haloes are randomly extracted from the \citet{tinker_2008} halo mass function above the halo mass cut-off. Since we were interested in the star formation histories of central galaxies only, we neglected any correction for satellite subhaloes (e.g. \citealt{behroozi_2013, fu_2022}).

    The mass accretion history of each dark matter halo and their merger trees are computed via the SatGen analytical code from \citet{jiang_2021}, based on the \citet{parkinson_2008} algorithm fitted to the outputs of the MultiDark Planck simulation. The SatGen merger trees, in addition to the evolution of the central subhaloes, also provide the accretion rate and mass evolution of the satellite subhaloes from their formation until infall, which we use to initialise satellite galaxies and black holes with stellar and black hole masses at the time of the merger by randomly and self-consistently extracting from the parent population of central galaxies of similar stellar and black mass in the mock (e.g. \citealt{hopkins_2009b, shankar_2014}). Hereafter, we refer to central subhaloes as central or parent haloes, and to satellite subhaloes simply as subhaloes.

    \subsection{Self-consistently assigning star formation rates to galaxies from the host halo accretion rates}\label{sec:decode_AM}

    Having set up the halo catalogues, we could grow galaxies by self-consistently following the accretion rate histories of their host haloes. More specifically, following \citetalias{fu_2025} and motivated by the results of state-of-the-art hydrodynamic simulations, we assign a SFR to a galaxy from a monotonic relation between SFR and HAR defined via the abundance matching relation (see Equation 37 in \citealt{aversa_2015})
    \begin{equation}\label{eq:aversa_AM}
    \begin{split}
        \int_{\log \dot{M}_{\rm \star, SFR}}^{+\infty} \phi( \log \dot{M}_{\rm \star, SFR}', z) \mathrm d \log \dot{M}_{\rm \star, SFR}' = & \\
        \int_{-\infty}^{+\infty} \frac{1}{2} \mathrm{erfc} \Bigg\{ \frac{\log \dot{M}_{\rm h} (\dot{M}_{\rm \star, SFR}) - \log \dot{M}_{\rm h}' }{\sqrt{2} \Tilde{\sigma}_{\log \dot{M}_{\rm \star, SFR}}} \Bigg\} \cdot & \phi( \log \dot{M}_{\rm h}', z) \mathrm d \log \dot{M}_{\rm h}' \; ,
    \end{split}
    \end{equation}
    where $\Tilde{\sigma}_{\log \dot{M}_{\rm \star,SFR}} = \sigma_{\log \dot{M}_{\rm \star,SFR}} / \mu $. Here, $\sigma_{\log \dot{M}_{\rm \star,SFR}}$ is the Gaussian scatter in SFR at fixed HAR, and $\mu = \mathrm d \log \dot{M}_{\rm \star,SFR} / \mathrm d \log \dot{M}_{\rm h}$ is the derivative of the SFR with respect to the HAR. This recipe provides an efficient tool to compute the SFR-HAR relation numerically with only a few inputs, the HAR and SFR distributions, and the scatter in $\log_{10} ({\rm SFR})$ at fixed $\log_{10} ({\rm HAR})$, without any pre-defined analytic fitting formula which requires a heavy Monte Carlo Markov chain (MCMC) exploration and the introduction of free parameters in the model. \citetalias{fu_2025} demonstrated that also in the TNG100 simulation the monotonic mean correlation between SFR and HAR can be reproduced via abundance matching between the SFR and HAR functions predicted by the simulation. We set the scatter in SFR at fixed HAR to $0.4$ dex, as suggested by cosmological simulation although our results do not alter within reasonable values of $0.3-0.5$ dex, as done in \citetalias{fu_2025}. The SFR-HAR relation computed from Equation (\ref{eq:aversa_AM}) is then used to assign galaxies with a SFR following the evolution of their host dark matter haloes.

    The two main inputs in Equation (\ref{eq:aversa_AM}) are the theoretically predicted HAR function $\phi( \log \dot{M}_{\rm h}, z)$ - from the outputs of N-body simulations - and the observational SFR function $\phi( \log \dot{M}_{\rm \star, SFR}, z)$. For the former, as done in \citetalias{fu_2025}, we make use of the HAR function computed from the mock catalogue at each redshift described in Sect. \ref{sec:decode_DM_halo_cat}. We note that our HAR function is consistent with the one calculated from the TNG simulation at all redshifts of interest. For the latter, we combine the most recent UV and IR data at multiple redshifts and compute a fitting formula to the SFR function, which we show in Sect. \ref{sec:res_sfr_har}. Following \citet{saunders_1990}, we choose a modified Schechter function to fit the SFR function data given by
    \begin{equation}\label{eq:saunders}
    \begin{split}
        \phi(\dot{M}_{\rm \star,SFR}) \mathrm{d} & \log \dot{M}_{\rm \star,SFR} = \phi^\star \bigg( \frac{\dot{M}_{\rm \star,SFR}}{\psi_{\rm \star,SFR}} \bigg)^{1-\alpha} \\
        \exp & \bigg[ - \frac{1}{2\sigma^2} \log^2_{10} \bigg( 1 + \frac{\dot{M}_{\rm \star,SFR}}{\psi_{\rm \star,SFR}} \bigg) \bigg] \mathrm{d} \log \dot{M}_{\rm \star,SFR} \;\;\; .
    \end{split}
    \end{equation}
    The Saunders fitting formula depends on the following free parameters $(\phi^\star, \psi_{\rm \star,SFR}, \alpha, \sigma )$. In particular, $\phi^\star$ regulates the normalisation of the distribution, $\psi_{\rm \star,SFR}$ the position of the knee, $\alpha$ and $\sigma$ the slopes of the faint and bright ends, respectively. We show the best-fitting values to these parameters at each redshift in Appendix \ref{app:fit_sfrf}.

    After assigning SFRs to haloes along their progenitor histories, SFRs are then integrated across cosmic time, to yield galaxy stellar masses, assuming that a fraction of stellar mass returns to the interstellar medium with loss rate given by $1- \mathcal{R}$, with $\mathcal{R}=0.45$ (e.g. \citealt{reimers_1975, renzini_1988, girardi_2000, pietrinferni_2004}). We start our simulation at redshift $z\simeq 6$, where we assume that all our galaxies are star-forming. This is a good approximation as the fraction of quenched galaxies at $z\sim6$ is relatively low. Indeed, recent results from the ASTRODEEP-JWST photometric catalogues reported a number density of $\sim 6.1\times 10^{-7} \, {\rm Mpc}^{-3}$ for quenched galaxies above $M_\star \gtrsim 10^{10.5} \,M_\odot$ at $5<z<7$ (\citealt{merlin_2025}) and the integral of the total stellar mass function (e.g. \citealt{weibel_2024, zalesky_2025}) produces a number density of $\sim 10^{-5} \, {\rm Mpc}^{-3}$ above the same stellar mass, yielding to a total quenched galaxies percentage of less than $\lesssim 6\%$ (see also \citealt{yang_2025}). We note that these galaxies contribute to the stellar mass growth of the nearby massive galaxies via major mergers producing the stellar population with old ages.

    \subsection{Merging galaxies}\label{sec:decode_mergers}

    Galaxy mergers are added by assigning the satellite galaxies to the dark matter subhaloes at infall by following the typical stellar mass growth of the central galaxies characterised by the same host halo mass, yielding the same results as using the mean stellar mass-halo mass (SMHM) relation. 
    As described in Sect. \ref{sec:decode_DM_halo_cat}, all parent haloes in our dark matter catalogue contain the mass growth of all the subhaloes that have ever accreted onto them. These subhaloes were not part of any halo system before the infall and were accreting mass as central haloes. We safely assume that satellite galaxies are frozen in stellar mass after infall. In fact, in \citet{fu_2024} (see also \citealt{grylls_2019}) we showed that on the assumption that stellar stripping and star formation balance out after infall, maintaining a nearly constant stellar mass, provides, at least at $M_\star \gtrsim 10^{10} \, M_\odot$, an excellent match to the local stellar mass function of central and (surviving) satellites as measured in the Sloan Digital Sky Survey. To each infalling satellite we assign a time delay following the merger timescales described in Sect. 3.4 of \citet{fu_2022} between the halo-halo merger and galaxy-galaxy merger. The latter timescales have been extracted from the Millennium N-body simulation and were tweaked by \citet{fu_2022} to reproduce the number densities of surviving subhaloes at each redshift of both the Millennium and SatGen. Galaxies that have not merged yet are labelled as surviving satellites at each redshift. Throughout, all the results shown refer to central galaxies, unless otherwise specified.

    \begin{figure*}
        \centering
        \includegraphics[width=\textwidth]{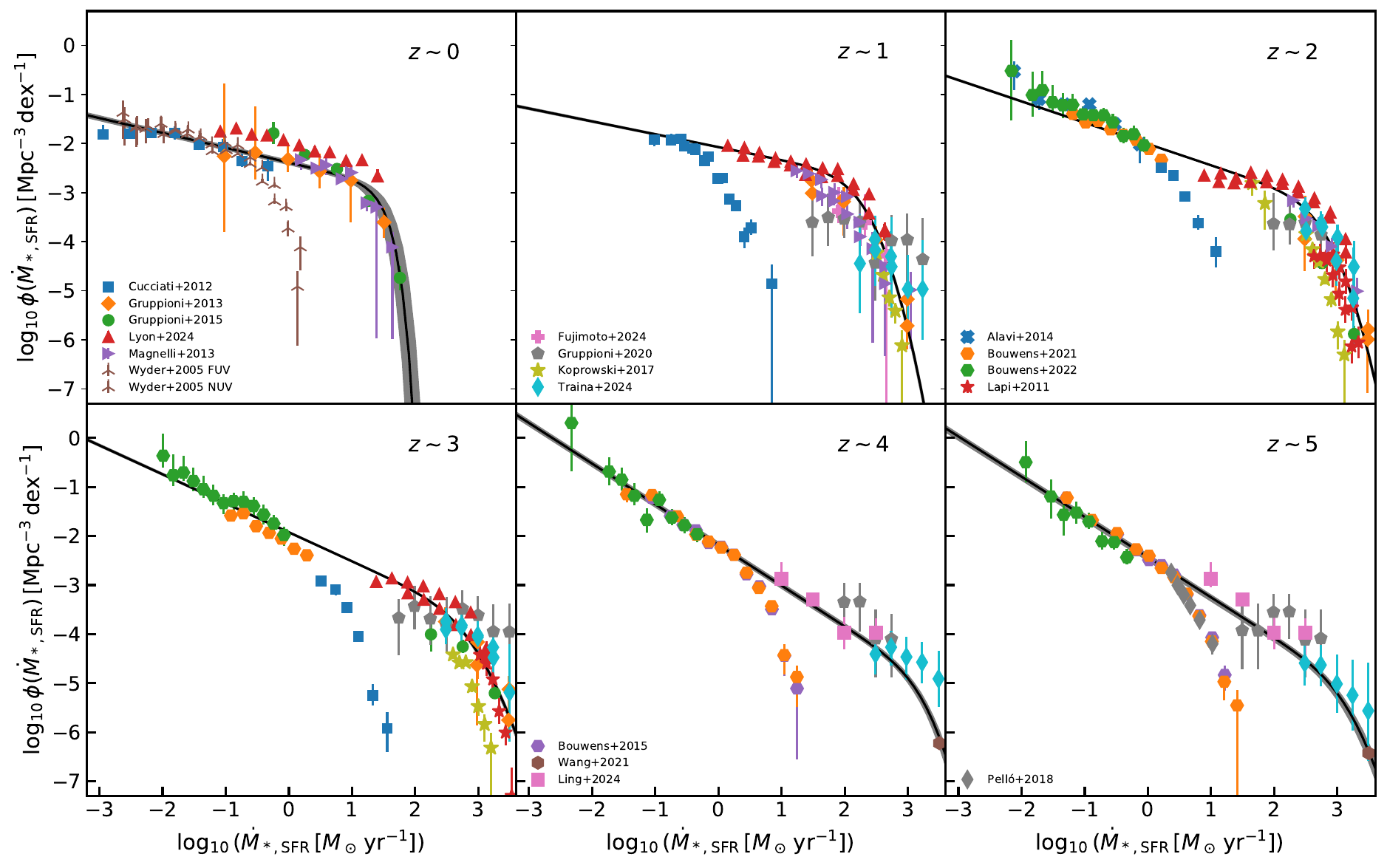}
        \caption{Star formation rate function at redshifts $z \sim 0$, $1$, $2$, $3$, $4$, and $5$. The data points with error bars include the data from VVDS (blue squares; \citealt{cucciati_2012}), Herschel PEP/HerMES (orange rhombuses; \citealt{gruppioni_2013}), PEP and HerMES (green circles; \citealt{gruppioni_2015}), CIGALE (red triangles; \citealt{lyon_2024}), Herschel-PACS (purple triangles; \citealt{magnelli_2013}), GALEX (brown arrows; \citealt{wyder_2005}), ALCS (pink pluses; \citealt{fujimoto_2024}), ALMA-ALPINE (grey pentagons; \citealt{gruppioni_2020}), SCUBA-2 (yellow stars; \citealt{koprowski_2017}), A$^3$COSMOS (cyan rhombuses; \citealt{traina_2024}), HST (blue crosses, purple, orange and green hexagons; \citealt{alavi_2014, bouwens_2015, bouwens_2021, bouwens_2022}), Herschel-ATLAS (red stars; \citealt{lapi_2011}), Herschel+LOFAR (brown hexagons; \citealt{wang_2021}), JWST-CEERS (pink squares; \citealt{ling_2024}) and WUDS (grey rhombuses; \citealt{pello_2018}). The black solid lines and shaded areas show our fit to Equation (\ref{eq:saunders}) and $1\sigma$ uncertainty.}
        \label{fg:phi_sfr_data}
    \end{figure*}

    \subsection{Growing supermassive black holes within DECODE}\label{sec:grow_SMBHs}

    To determine the accretion rate $\dot{M}_{\rm BH}(M_\star,z)$ competing to each SMBH living in a host galaxy of stellar mass $M_\star$ at redshift $z$, we made use of the empirical determination from \citet{bongiorno_2016} of the $P(\lambda| M_\star,z)$, measured using X-ray selected sources from the XMM-COSMOS survey. $P(\lambda| M_\star,z)$ is defined as the number density of active galaxies at a given stellar mass accreting at $\lambda = L_{\rm X} / M_\star$ in units of ${\rm Mpc}^{-3} \, {\rm dex}^{-1}$. The $P(\lambda| M_\star,z)$ includes information on both the duty cycle $U(M_\star,z)$, i.e. the fraction of galaxy of stellar mass $M_\star$ hosting a SMBH accretion above $\lambda_{\rm min}$, and the (normalised) probability $p(\lambda| M_\star,z)$ that a galaxy of mass $M_\star$ at $z$ hosts a SMBH accreting at the $\lambda$ rate. The former, can be simply computed by integrating $P(\lambda| M_\star,z)$ at fixed $M_\star$ above the minimum luminosity of the survey at $z$. The latter $p$ distribution is instead determined by normalising $P(\lambda)$ to unity at fixed $M_\star$.

    At each redshift we predefined the fraction of active galaxies (i.e. hosting an accreting AGN) via the input duty cycle. To each active galaxy, with given stellar mass and redshift, we then assigned a $\lambda$ randomly extracted from the $P(\lambda)$ distribution. When moving forward in time, we shut down the AGN activity by updating the fraction of active galaxies via the duty cycle. Since the AGN is a random event which occurs stochastically for a quasar lifetime scale, as supported by results from several works (e.g. \citealt{martini_2004, hopkins_2005, shankar_2009, yuan_2018, khrikin_2021, shen_2021}), we resampled the AGN population after each 100 Myr. The normalised luminosity $\lambda$ multiplied by the galaxy stellar mass will provide the luminosity of the black hole in the X-ray band. In order to estimate the mass accretion rate of the black hole we converted the X-ray luminosity into bolometric luminosity $L_{\rm bol}$ via the \citet{duras_2020} conversion factor. Our results do not change if we adopted any other bolometric correction (e.g. \citealt{marconi_2004, lusso_2012, yang_2018}). We then converted $L_{\rm bol}$ into BHAR using the $\dot{M}_{\rm BH} = (1 - \epsilon - \epsilon_{\rm kin}) L_{\rm bol} / \epsilon c^2$ relation, where we assumed a radiative efficiency of $\epsilon=0.1$ and kinetic efficiency due to winds and jets of $\epsilon_{\rm kin} = 0.15$\footnote{Varying the values of $\epsilon_{\rm kin}$ within reasonable ranges of $0.003-0.2$ alters the resulting black holes masses by less than $0.05$ dex.} (e.g. \citealt{shankar_2008, shankar_2020}).

    Equipped with the accretion rates of the black holes, we then simply integrated them across cosmic time to compute the mass growth of the SMBHs. Starting the simulations at $z\sim 6$, we experiment different ways of initialising the black hole mass. As shown in several works, the choice of the black hole seeds does not significantly impact the mass growth of the SMBHs at redshift $z\lesssim3$ (e.g. \citealt{zou_2024}). In particular, we explored a black holes seeding via the observed $M_{\rm BH}-M_\star$ relations from \citet{reines_volonteri_2015} and \citet{li_2025}. We also studied the possibility of initialising our black holes with a seed of mass $M_{\rm BH,seed}\sim 10^5 \, M_\odot$ when the halo mass reaches $M_{\rm h} \sim 10^{10}\, M_\odot$, similarly to what is assumed in hydrodynamical simulations and analytical models (e.g. \citealt{schaye_2015, rosas-guevara_2016, weinberger_2018, dave_2019}). We find that by seeding SMBHs via any of the approaches described above does not noticeably alter the growth histories of SMBHs, at least at $z<3$. In what follows, we choose the seeding of $M_{\rm BH,seed}\sim 10^5 \, M_\odot$ as a reference, but also compare with the predictions from the other seeding models.

    In addition to gas accretion, we allowed black holes to grow via mergers with other black holes residing inside satellite galaxies, to which we assigned the corresponding black hole mass from corresponding the satellite's stellar mass from the $M_{\rm BH}-M_\star$ distribution of the central counterparts. Following the analytic work of \citet{izquierdo-villalba_2020} based on the results from hydrodynamical simulations, we applied a time delay between the mergers of the host galaxies and the central SMBHs via the following fitting formula (Equation 29 of \citealt{izquierdo-villalba_2020})
    \begin{equation}\label{eq:tau_merge_BH}
        \tau_{\rm BH,delay} = 0.01 \bigg( \frac{0.1}{q} \bigg) \bigg( \frac{0.3}{f_{\rm gas}} \bigg) \mathcal{F}(e) \; {\rm Gyr} \, ,
    \end{equation}
    where $q$ is the mass ratio between the two merging black holes, $f_{\rm gas}$ is the gas fraction of the host galaxy assigned via input empirical $M_{\rm gas}-{\rm SFR}$ relations (e.g. \citealt{stewart_2009, santini_2014}), and $\mathcal{F}(e)$ is a factor that accounts for the dependence on the eccentricity which for simplicity was assumed to be equal to $1$\footnote{Since we find that the contribution of mergers is overall negligible in the total mass growth of SMBHs, we do not investigate the effect of varying this input parameter.}. We note that the black hole mergers are sensible to the assumed timescale and evolution after infall. Although the effect is minimal, we stress that assuming all satellite black holes are inactive is extreme, as the $P(\lambda)$ is calibrated on all galaxies (including both centrals and satellites) and AGN clustering analysis showed that some satellites are active (e.g. \citealt{shankar_2020NatAs, shen_2020}). However, the aforementioned parameters should not affect our results, since the overall mass growth is dominated by the gas accretion channel. 

    \subsection{Quenching galaxies in our semi-empirical framework}\label{sec:decode_quench}

    We quenched galaxies when their central SMBHs surpass a mass threshold set by a certain scaling relation. In this work, we take as a reference the black hole mass-stellar velocity dispersion $M_{\rm BH}-\sigma_\star$ relation which, as discussed by, e.g. \citet{shankar_2025} (and references therein), presents the strongest residuals, at least within the sample of local inactive SMBHs. In addition, several groups have shown that it is quite stable with redshift and all AGN samples align on similar $M_{\rm BH}-\sigma_\star$ relations (e.g. \citealt{shankar_2009, marsden_2022, maiolino_2024} and references therein).

    To assign a stellar velocity dispersion $\sigma_\star$ to a galaxy we follow the size, stellar mass, and redshift dependent analytic fits derived by \citet{marsden_2022}, based on $\sigma_\star$ extracted from a Jeans equation, inclusive of a S\'ersic profile for the stellar component and a Navarro–Frenk–White (NFW) profile for the dark matter component. We here choose an aperture equal to the effective radius. We note that none of our main results depend on the assumed specific aperture for $\sigma_\star$. When quenching galaxies, expanding on the modelling put forward by \citetalias{fu_2025}, we include a delayed decay (e.g. \citealt{daddi_2022}) to shut down the SFR via the following exponential formula
    \begin{equation}\label{eq:tau_quench}
        \dot{M}_{\rm \star,SFR} (t) = \dot{M}_{\rm \star,SFR} (t_{\rm q}) \cdot e^{ - (t - t_{\rm q}) / \tau } \;\;\; {\rm for} \;\;\; t>t_{\rm q} \; ,
    \end{equation}
    where $t_{\rm q}$ is the cosmic time when the quenching process starts, defined as the time when the galaxy crosses for the first time the $M_{\rm BH}-\sigma_\star$ relation, and $\tau$ is the quenching timescale. We include a dispersion of $0.3$ dex around the mean $M_{\rm BH}-\sigma_\star$ quenching threshold\footnote{We use, as quenching threshold, a random number assigned following a Gaussian distribution with $\sigma=0.3$ and centred in the mean $M_{\rm BH}-\sigma_\star$ relation.}, as suggested by the observations. We adopt as a reference value for the quenching timescale $\tau=1$ Gyr, which reflects what inferred from SED-derived star formation histories (e.g. \citealt{bellstedt_2020, bertemes_2023, davies_2025, wan_2025}), noticing that allowing $\tau$ to vary within $0.5-2$ Gyr has a small impact on the predicted fractions of quenched galaxies. We then define the quenched galaxies as those galaxies with specific star formation rate (sSFR) below $\dot{M}_{\rm \star,SFR} / M_\star \lesssim 10^{-11} \, {\rm yr}^{-1}$.

    In order to study the quenching due to the black hole feedback, we proceed as follows. We assumed that, whenever a black hole crosses the $M_{\rm BH}-\sigma_\star$ relation, the central SMBH becomes massive and bright enough to heat or eject the cold gas reservoir available for star formation. In other words, we assumed that the final position of the black hole on the $M_{\rm BH}-\sigma_\star$ relation is linked to the quenching phase, as suggested by several theoretical and observational groups (e.g.  \citealt{granato_2004, king_2015, martin_navarro_2018}). We have explored the impact on our results of adopting different reference $M_{\rm BH}-\sigma_\star$ relations from the literature (e.g. \citealt{kormendy_ho_2013, mcconnell_2013, woo_2015, shankar_2016, de_nicola_2019, caglar_2020}) as input in our model, finding that they all yield similar results, with the fraction of predicted quenched galaxies varying by less than $10\%$. In what follows, we adopt as a reference the $M_{\rm BH}-\sigma_\star$ scaling relation calibrated by \citet{de_nicola_2019}, as it falls in the mean region of the observed relations in the literature.

\section{Results}\label{sec:results}

In this section, we present our main results. In particular we focus on the 1) star formation histories characterising the host galaxies, derived from abundance matching; 2) the predicted mass growth histories of SMBHs; 3) the resulting fraction of quenched galaxies; and 4) the related derived scaling relations with stellar mass and SFR to compare with independent datasets. We show that our approach can self-consistently and simultaneously reproduce all these distinct observables whilst providing specific predictions on the quenching rate of central galaxies.

    \subsection{The star formation rate-halo accretion rate relation}\label{sec:res_sfr_har}

    We start by presenting the SFR function and SFR-HAR relation used as inputs in our semi-empirical framework. Figure \ref{fg:phi_sfr_data} shows the SFR number densities at redshifts $z=0-5$ that we use in our abundance matching. In particular, we combine the latest determination of the galaxy LF from different datasets both in the ultraviolet (UV) and infrared (IR) such as HST (\citealt{bouwens_2015, bouwens_2021, bouwens_2022}), UltraVISTA+UDS (\citealt{bowler_2015}), WUDS (\citealt{pello_2018}), ALMA-ALPINE (\citealt{gruppioni_2020}), Herschel+LOFAR (\citealt{wang_2021}), JWST-CEERS (\citealt{ling_2024}) and A$^3$COSMOS (\citealt{traina_2024}). We convert luminosities into SFRs via the \citet{madau_dickinson_2014} relation, ${\rm SFR} = \mathcal{K} \cdot L$, with $\mathcal{K} = 2.5 \times 10^{-10} M_\odot {\rm yr}^{-1} L^{-1}_\odot$ in the UV band and $\mathcal{K} = 1.73 \times 10^{-10} M_\odot {\rm yr}^{-1} L^{-1}_\odot$ in the IR band. We note that the \citet{madau_dickinson_2014} conversion is calibrated over a wide range of multi-wavelength data and is slightly sensitive to the choice of the spectral energy distribution, metallicity evolution, stellar evolution and IMF\footnote{We have applied the due conversion to the \citet{chabrier_2003} IMF.}. We fit the above data to Equation (\ref{eq:saunders}) at each redshift via MCMC chains. The black solid lines in Figure \ref{fg:phi_sfr_data} indicate the best-fitting curves for the observational data to the \citet{saunders_1990} formula (Equation \ref{eq:saunders}), with best-fitting parameters and corner plots of the fit reported in Appendix \ref{app:fit_sfrf}. Similarly to what found by other works (e.g. \citealt{sargent_2012, mancuso_2016, fujimoto_2024}) at redshift $z\lesssim3$, the observational SFR functions can be well described by a power law which drops towards the bright end and present a rapid evolution between $z=0$ to $z\sim1$, whilst maintaining a self-similar shape up to $z\sim6$. Such an abrupt change at low redshift, in the larger time gap between $z=1$ and $z=0$, may be attributed to the more marked emergence of quenched galaxies.

\begin{figure}
        \centering
        \includegraphics[width=\columnwidth]{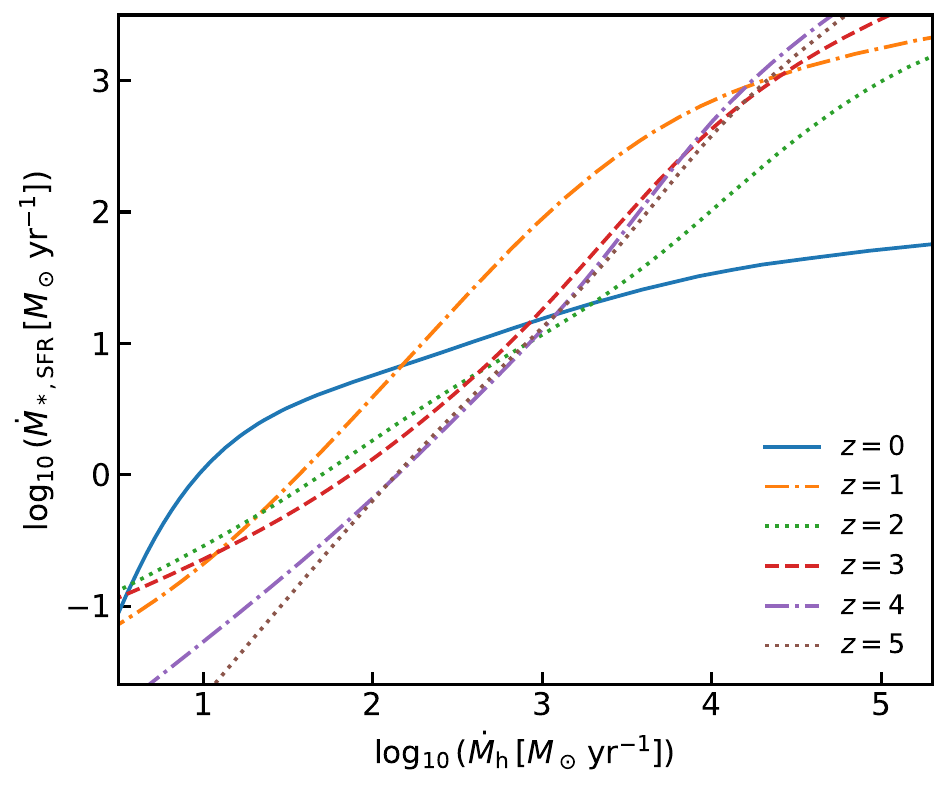}
        \caption{Star formation rate-halo accretion rate relation at redshifts $z = 0$, $1$, $2$, $3$ $4$ and $5$, from the abundance matching using as input the star formation rate function described in Sect. \ref{sec:res_sfr_har}.}
        \label{fg:sfr_har_z456}
    \end{figure}

    Figure \ref{fg:sfr_har_z456} shows the SFR-HAR relation at $z = 0$, $1$, $2$, $3$, $4$ and $5$ computed from abundance matching using the input SFR functions described above. The relation presents a log-linear function with slope $\sim 1.5$ broadly constant in redshift above $z\gtrsim1$. Instead, in the local Universe the relation follows a double power law shape, as we also showed in \citetalias{fu_2025}. We then use the above SFR-HAR relation to assign the SFRs following the HAR histories of the host dark matter haloes, and then integrate the SFRs across cosmic time.

    \subsection{Black hole mass accretion rates}\label{sec:BH_mass_acc}

    \begin{figure}
        \centering
        \includegraphics[width=\columnwidth]{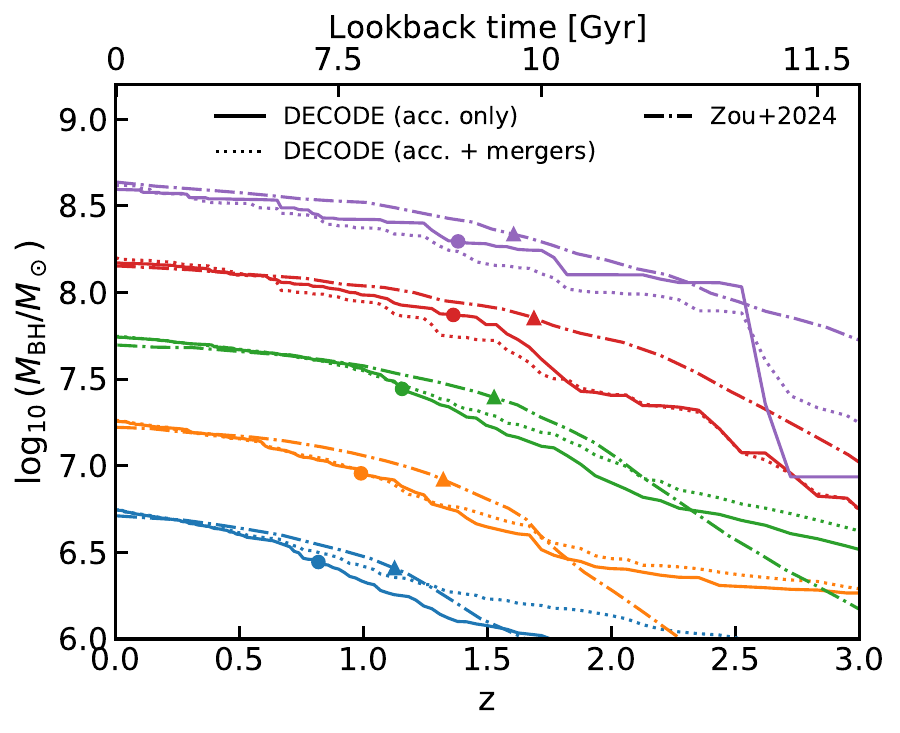}
        \caption{Average black hole mass accretion histories as predicted from \decode, compared to those predicted from \citet[][dash-dotted lines]{zou_2024}. \decode's results are shown both for the accretion-only scenario (solid lines) and the scenario with mergers (dotted lines). The filled circles and triangles show the redshift at which half of the black hole mass is formed. The different colours show the accretion tracks for different black mass bins at $z=0$: $\log_{10} (M_{\rm BH}/M_\odot) \simeq 6.75$ (blue lines), $\log_{10} (M_{\rm BH}/M_\odot) \simeq 7.25$ (orange lines), $\log_{10} (M_{\rm BH}/M_\odot) \simeq 7.75$ (green lines), $\log_{10} (M_{\rm BH}/M_\odot) \simeq 8.2$ (red lines), and $\log_{10} (M_{\rm BH}/M_\odot) \simeq 8.6$ (purple lines).}
        \label{fg:mbh_acc_tracks}
    \end{figure}

    On top of the galaxy stellar mass assemblies, we grew the central SMBHs by assigning a BHAR at each timestep along the evolution following the $\lambda$ distribution functions described in Sect. \ref{sec:grow_SMBHs}, and subsequently integrating the BHARs across cosmic time along the progenitor mass accretion tracks of the hosts to compute the black hole mass. As described in Sect. \ref{sec:grow_SMBHs}, black hole masses are initialised with seeds of mass $10^{5}\, M_\odot$ as soon as the host halo mass crosses the $10^{10} \, M_\odot$ threshold, although we checked we obtain similar results when we adopt a seeding from a $M_{\rm BH}-M_\star$ relation (e.g. \citealt{reines_volonteri_2015, li_2025}). We note that our generated SMBH mass growth histories are consistent with the AGN bolometric luminosity function (see Appendix \ref{app:gal_agn_abun}), and thus it preserves the So\l tan-energy argument (e.g. \citealt{soltan_1982, salucci_1999}), as expected, given that the input Eddington ratio distributions, when convolved with the galaxy stellar mass function, yield the observed bolometric AGN LF (e.g. \citealt{ueda_2014, aird_2015, bongiorno_2016}). We are also able to match the observed BHAR-stellar mass relation from X-ray analysis (Appendix \ref{app:Mbhdot_Mstar}), further validating the robustness of our model.

    \begin{figure}
        \centering
        \includegraphics[width=\columnwidth]{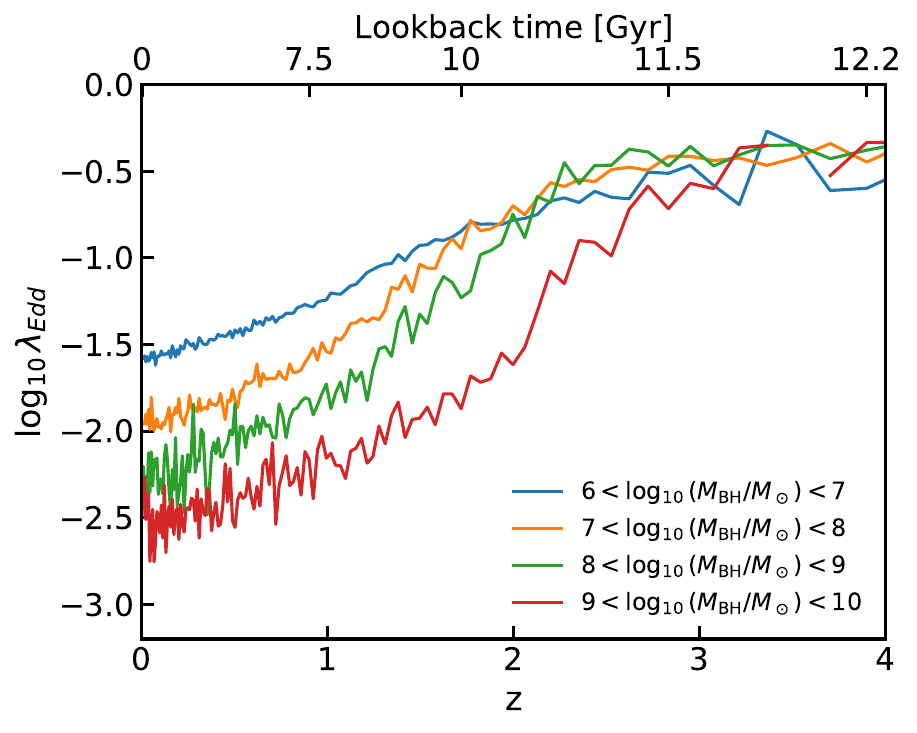}
        \caption{Average of the logarithm of the Eddington ratio, $\lambda_{\rm Edd}$, as a function of redshift for different black hole mass bins at redshift $z=0$, as labelled.}
        \label{fg:Eratio_evo}
    \end{figure}

    \begin{figure*}
        \centering
        \includegraphics[width=\columnwidth]{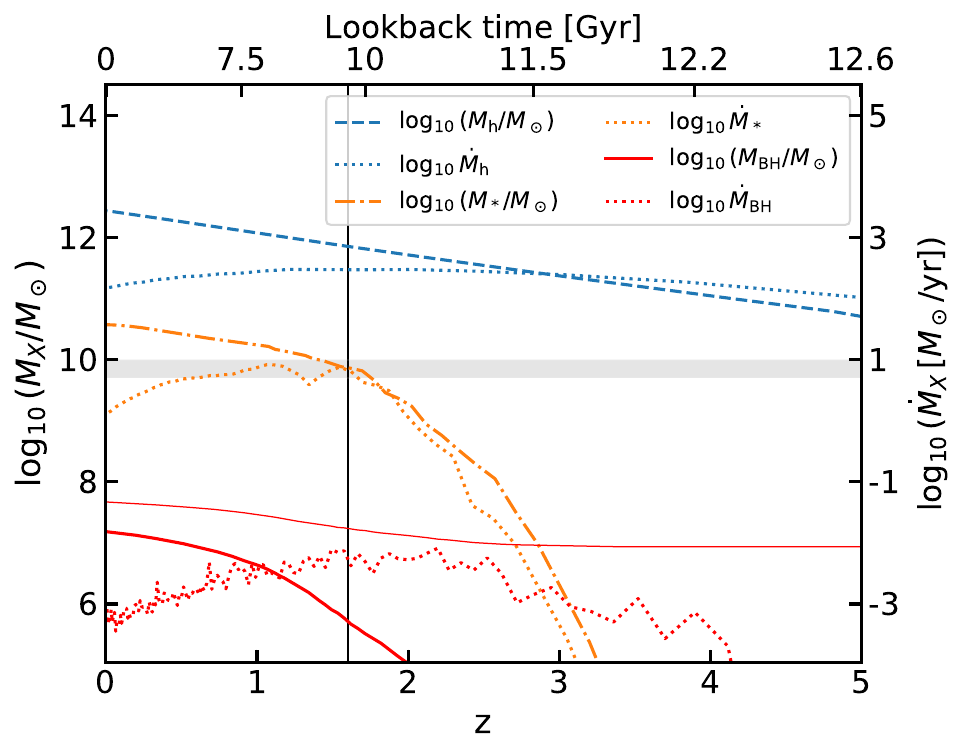}
        \includegraphics[width=\columnwidth]{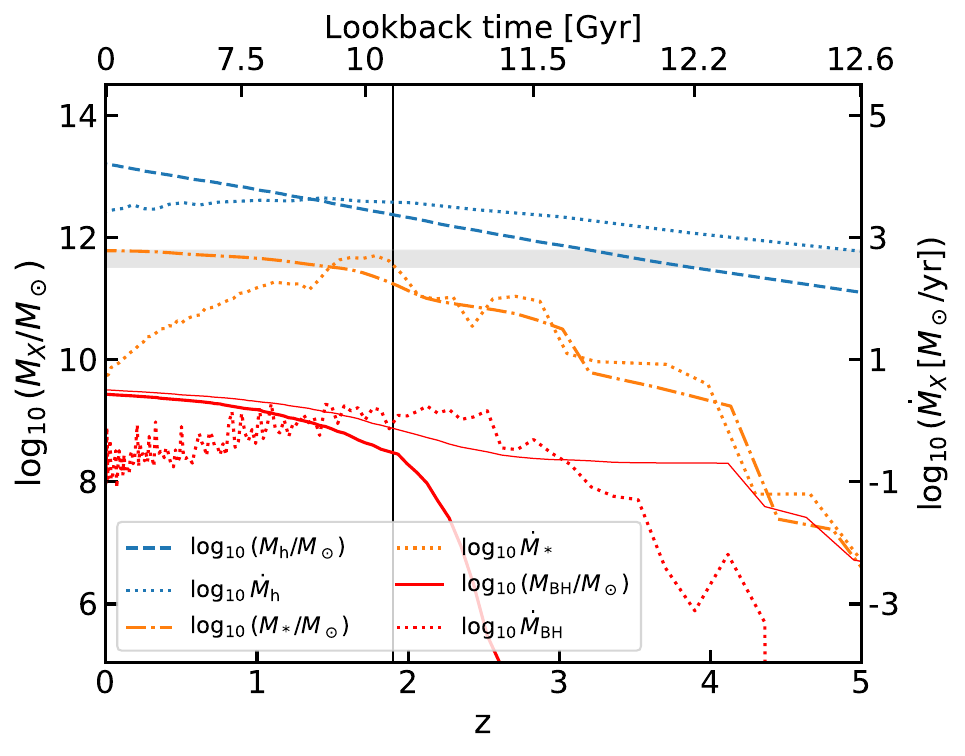}
        \hspace*{-1.15cm}
        \includegraphics[width=0.95\columnwidth]{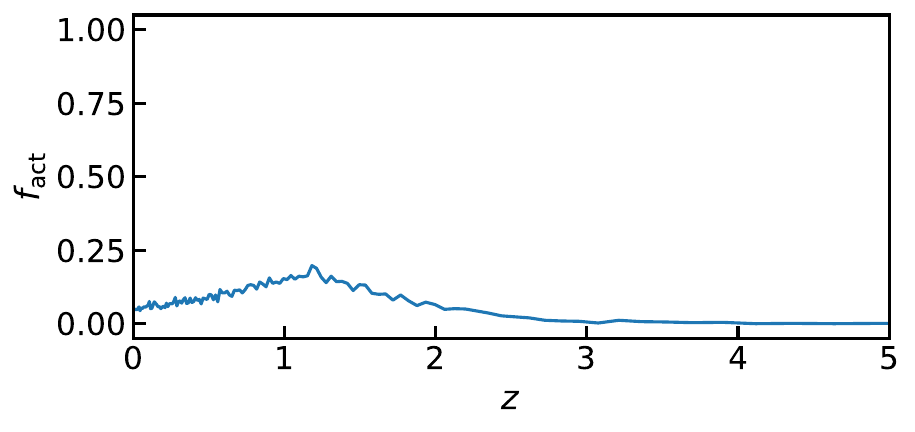}
        \hspace*{0.33cm}
        \includegraphics[width=0.95\columnwidth]{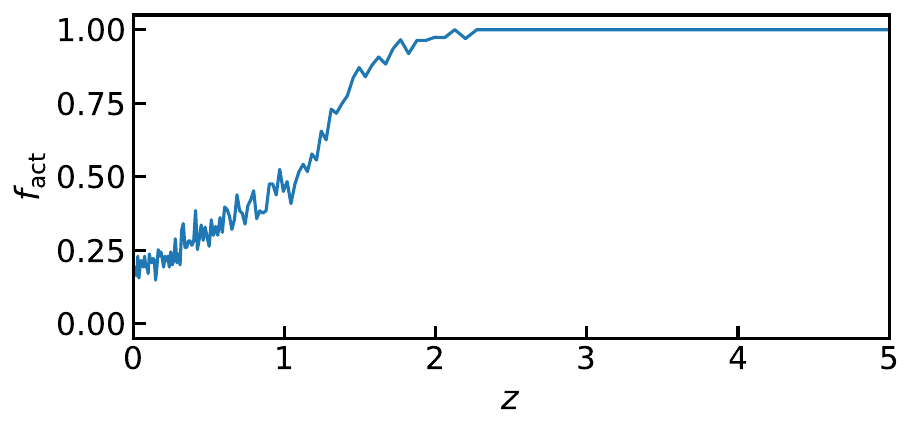}
        \caption{Upper panels: Mean evolution of galaxies in our catalogue selected with stellar mass of $M_\star \simeq 10^{10.5} \, M_\odot$ (left) and $M_\star > 10^{11.5} \, M_\odot$ (right) at redshift $z=0$ from \decode's catalogue. The blue dashed and dotted lines show the host dark matter halo mass assembly history and accretion rate. The orange dash-dotted and dotted lines show the galaxy stellar mass growth and star formation rate, respectively. The red solid and dotted lines show the black hole mass assembly and accretion rate, respectively. The thin red solid lines show the case where we use the JWST \citet{pacucci_2023} relation to seed the SMBHs. The left-hand side y-axis refers to the mass growth values and the right-hand side y-axis refers to the accretion rate values. The horizontal shaded area shows the $M_{\rm BH} - \sigma_\star$ quenching threshold with the 0.3 dex dispersion. The vertical black line shows the redshift at which the host halo reached the golden halo mass $M_{\rm h} \sim 10^{12}\, M_\odot$. Lower panels: Fraction of AGNs as a function of redshift for the same galaxy stellar mass bin as in the upper panel.}
        \label{fg:ex_mbh_growth}
    \end{figure*}

    Figure \ref{fg:mbh_acc_tracks} reports the mean black hole mass accretion tracks predicted by \decode for different $z=0$ black hole masses (solid lines). Dotted lines are the predicted black hole assembly tracks inclusive of mergers. We compare our predictions to those from \citet{zou_2024} obtained by assigning an BHAR to each mean galaxy assembly track from the mean $\dot{M}_{\rm BH}-M_\star$ relation derived from observations (dash-dotted lines). Our prediction are in very good agreement with those from \citet{zou_2024}, despite some differences in the underlying assumed galaxy growth histories and the fact that they use mean relations instead of single-source growth tracks as carried out here.
    It is interesting to note that all SMBHs of any relic mass reach $50\%$ of their final mass around $z\sim1-1.5$, with only a relatively mild downsizing, in agreement with earlier studies from e.g. \citet{shankar_2020} and \citet{zou_2024}. We also note that mergers tend to have a minimal impact on the total mass growth histories in black holes with $M_{\rm BH} < 10^9 M_\odot$ at $z=0$ (compare solid and dotted lines), although we expect some dependence on the assumed merger timescale (see Sect. \ref{sec:grow_SMBHs}).

    We now turn our attention to the time evolution of the Eddington ratio, defined as the ratio between the BHAR and the Eddington limit $\lambda_{\rm Edd} = \dot M_{\rm BH} / \dot M_{\rm Edd} $\footnote{We note that the observed Eddington ratio is typically defined as $L_{\rm bol} / L_{\rm Edd}$ and we checked that similar results are found when using the latter definition and it fits the data on available Eddington ratios.}, which contains information on the black hole mass growth efficiency. Figure \ref{fg:Eratio_evo} shows the mean Eddington ratio as a function of redshift for different black hole mass bins. All black holes show a steady decrease in (mean) Eddington ratio by at least an order of magnitude from $z\sim3$ to $z\sim0$, with the most massive black holes showing the largest decline, which represents a more evident sign of downsizing in terms of mass accretion rates. This behaviour in Eddington ratios is expected given that the input cumulative Eddington distribution function is declining with redshift. Our results in Figure \ref{fg:ex_mbh_growth} indicate that also the duty cycle, i.e. the fraction of black holes active above the minimum luminosity of current X-ray surveys, $L_{\rm bol} > 10^{42} \, {\rm erg/s}$, tends to decrease with cosmic time, especially for the most massive black holes, in line with continuity equation results (e.g. \citealt{shankar_2013, aversa_2015}).

    \subsection{Black hole-galaxy co-evolution}\label{sec:BH_gal_coevo}

    Galaxy evolution shows a sharp change in behaviour around the so-called golden halo mass ($M_{\rm h} \sim 10^{12} \, M_\odot$). In particular, below this threshold galaxies are found to be mostly blue with their stellar mass growth dominated by star formation, whereas above this threshold their stellar mass growths are found to be mostly dominated by mergers. Numerical simulations including AGN feedback are capable of broadly capturing this galaxy sequence (although a clear consensus has yet to be reached, e.g. \citealt{schaye_2015, choi_2017, pillepich_2018_MstarContent, dave_2019, zhu_2023}), as also implemented in this work and described in Sect. \ref{sec:decode_quench}. In this section, we investigate the effect of black hole quenching, its possible connection to the host halo, and the resulting co-evolution with the host galaxy.

        \subsubsection{Black hole energy feedback and star formation histories}\label{sec:BHfeedback_SFH}

        The upper panels of Figure \ref{fg:ex_mbh_growth} show the mean evolutionary tracks of the haloes from \decode's catalogue selected from two central galaxy stellar mass bins, $M_\star \simeq 10^{10.5} \, M_\odot$ and above $M_\star > 10^{11.5} \, M_\odot$ at redshift $z=0$. In particular, we show the halo mass assembly track (blue dashed and dotted lines), along with the galaxy star formation history (orange dash-dotted and dotted lines) and black hole accretion history (red solid and dotted lines). By design, the SFR starts decreasing when the black hole mass reaches the $M_{\rm BH} - \sigma_\star$ limit (denoted by the horizontal shaded areas) and, with a minimal time delay, the SMBH accretion rate also starts decreasing, which is remarkable. Interestingly, while the SFR's decrease is an input of the model, the decline in SMBH accretion rate along the SFR is a genuine prediction of the model, as in our method the SMBH continues accreting at the rates predicted by the $P(\lambda)$ at all times, even after reaching the $M_{\rm BH} - \sigma_\star$ relations.  The relatively small delay of $0.7$ Gyr between the two peaks is a simple consequence of the delay model given in Equation (\ref{eq:tau_quench}). In addition, we also find that the epoch at which the black hole mass approaches the $M_{\rm BH} - \sigma_\star$ limit, is also the epoch when the host halo reaches the golden mass limit of $M_{\rm h} \sim 10^{12} \, M_\odot$. In other words, at least at this level of the analysis, the SMBH quenching and halo quenching would appear quite degenerate, as we also further discuss below. The initial stages of the black hole mass growth are sensitive to the seeding (thin and thick red solid lines). The black hole mass is higher, for both the mass bins shown in Figure \ref{fg:ex_mbh_growth}, when using the JWST (\citealt{pacucci_2023}) $M_{\rm BH}-M_\star$ relation with respect to the case of adopting the halo mass threshold seeding or \citet{reines_volonteri_2015} relation, although the evolutionary histories are nearly identical at late times, below $z<2$. The difference in black hole mass due to the seeding choice is more pronounced in the lower mass black holes above $z\gtrsim 2$.

        We note that in our modelling, the quenching resembles a quasar-mode, impulsive feedback scenario, which permanently quenches the host galaxy. The choice of this prescription is motivated by scenarios in which luminous AGNs eject or heat the galaxy cold gas reservoir on short timescales. We do not allow for any reactivation of the galaxy at later times, which could mimic the underlying effect of a longer-paced kinetic-mode feedback (such as jets; e.g. \citealt{harrison_2017, blandford_2019, torrey_2020}). Quenching due to kinetic-mode feedback is characterised by alternate episodes of heating of the circumgalactic medium, instead of a single quick ejection of gas, and allows for the possibility of galaxy reactivation or rejuvenation. Nevertheless, the integrated effect of a kinetic and/or late-time quasar-mode feedback is still to prevent further significant star formation in the galaxy which remains largely quenched (e.g. \citealt{granato_2004, croton_2006, fontanot_2020}).

\begin{figure}
            \centering
            \includegraphics[width=\columnwidth]{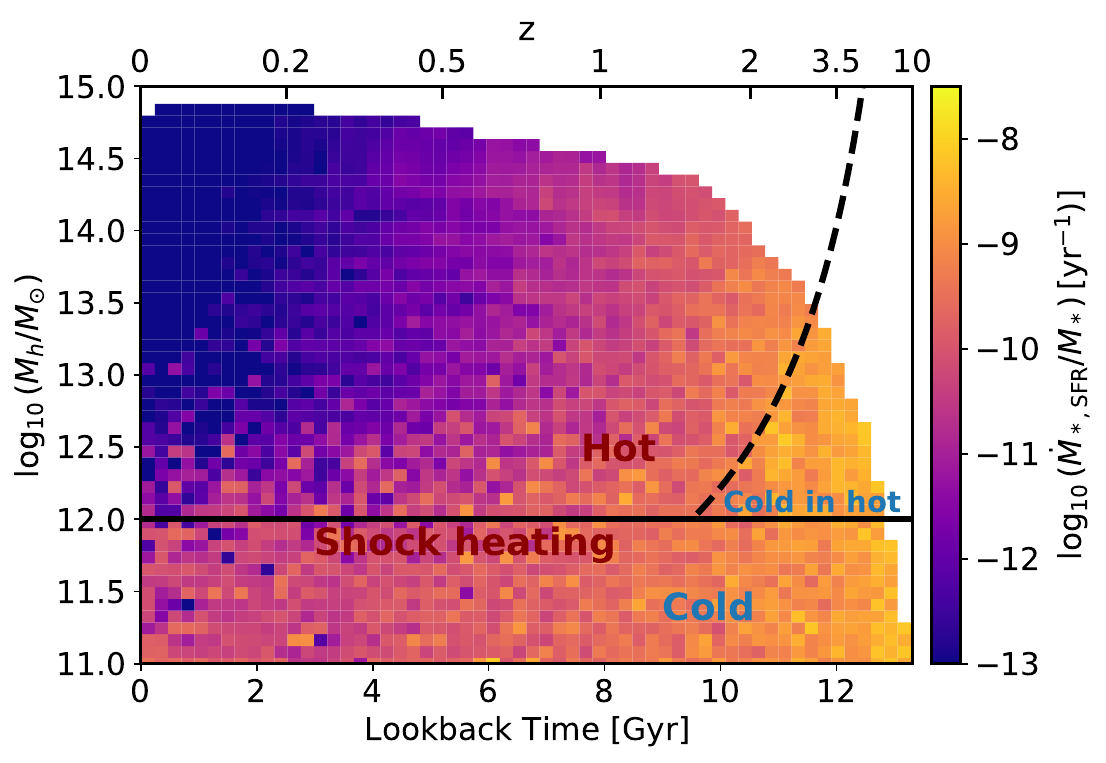}
            \includegraphics[width=\columnwidth]{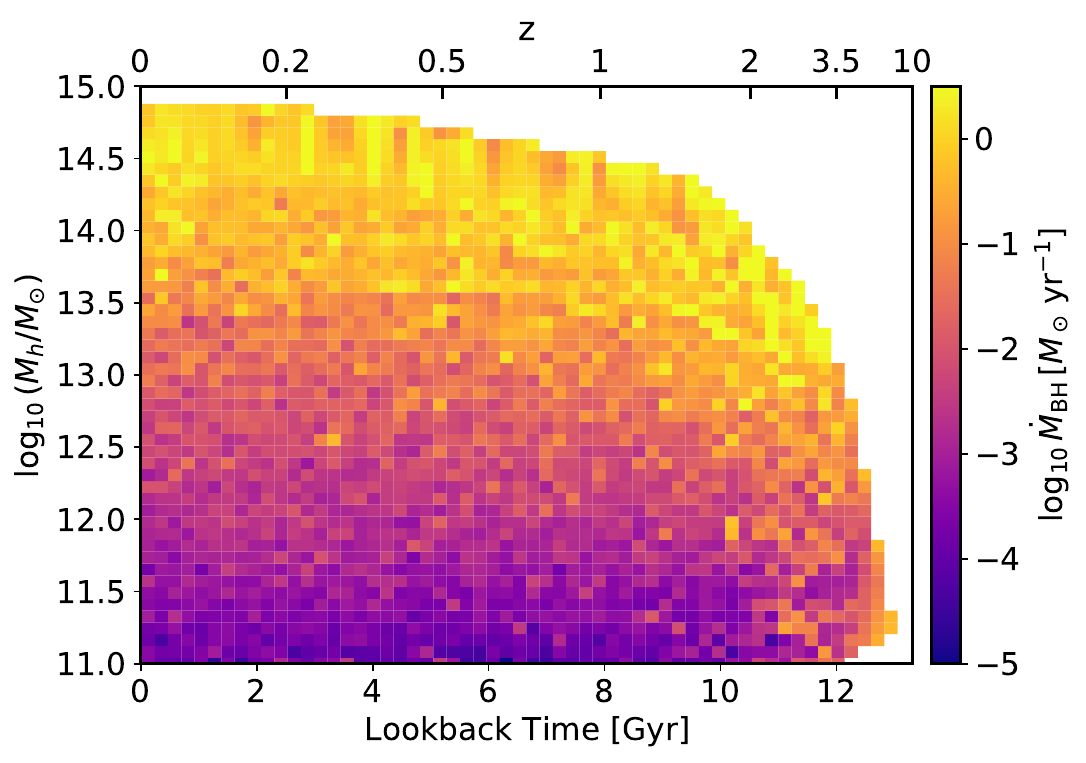}
            \caption{Upper panel: Specific star formation rate as a function of halo mass and redshift. The black (horizontal line) solid and dashed lines (with slope 1 and crossing $M_{\rm h}=10^{12}\, M_\odot$ at $z=1.5$) show the golden halo mass threshold that distinguishes the hot, cold and cold-in-hot modes, respectively (see \citealt{dekel_2006} and \citealt{fu_2025}). Lower panel: Same as upper panel but for the black hole accretion rate.}
            \label{fg:mh_acc_sSFR_BHAR}
        \end{figure}

\begin{figure}
            \centering
            \includegraphics[width=\columnwidth]{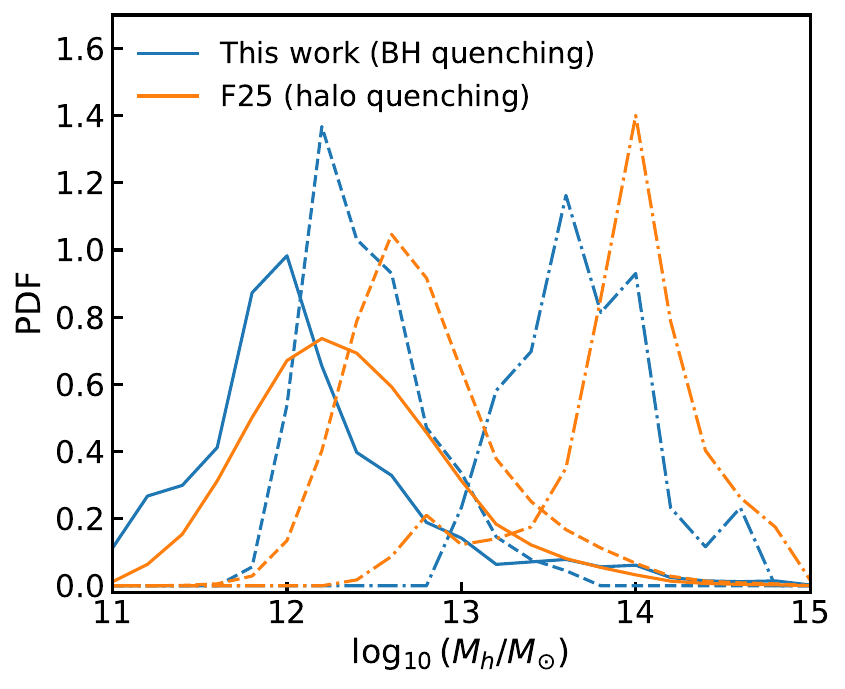}
            \caption{Distribution, normalised to unity, of the halo masses for the black hole quenching and halo quenching (\citealt{fu_2025}) models. The distributions are shown for all galaxies (solid lines), $10^{10.5} \, M_\odot < M_\star< 10^{11}\, M_\odot$ (dashed lines) and $M_\star > 10^{11.5} \, M_\odot$ (dash-dotted lines).}
            \label{fg:mhalo_dist}
        \end{figure}

        We more comprehensively view the connection between BHAR, SFR, and halo growth in Figure \ref{fg:mh_acc_sSFR_BHAR}, which shows the galaxy sSFR (upper panel) and BHAR (lower panel) as a function of the halo mass accretion history for all galaxies and black holes in our catalogue. We find that below $z \lesssim 1.5$ most galaxies living in haloes above $M_{\rm h} \gtrsim 10^{12} \, M_\odot$ are quenched, as we also showed in \citetalias{fu_2025}. Interestingly, an empirical quenching recipe given by the $M_{\rm BH}-\sigma_\star$ relation as employed in this paper, leads to qualitatively similar results to what suggested by the halo quenching, the two processes are quite degenerate between each other. In fact, as shown in the lower panel of Figure \ref{fg:mh_acc_sSFR_BHAR}, high values of the BHAR are favoured in more massive haloes. This result is also in line with the results from AGN clustering (e.g. \citealt{georgakakis_2019, allevato_2021}), which suggests that relatively luminous AGNs are usually hosted in dark matter haloes with mass above $M_{\rm h} \gtrsim 10^{12} \, M_\odot$. Figure \ref{fg:mhalo_dist} compares the distributions of halo masses for the halo quenching model (\citetalias{fu_2025}) and the black hole quenching model from this work, for different bins of stellar mass at $z=0$ (all galaxies, $10^{10.5} \, M_\odot < M_\star< 10^{11}\, M_\odot$ and $M_\star > 10^{11.5} \, M_\odot$). We find that although qualitatively similar, the two predicted halo distributions are somewhat quantitatively different, with an overall average offset of $\sim 0.3$ dex. The galaxies in the SMBH quenching model tend to reside in host halo masses systematically lower by a factor of $\sim2.5$ due to the slightly different output stellar mass-halo mass relations in the two models. This prediction can be tested via the clustering of AGN. Some preliminary results by \citet{leauthaud_2015} (Figure 8 therein) indeed show that galaxies of the same stellar mass tend to reside in haloes more aligned with the SMBH quenching model, but the systematics in all mass scales involved in this analysis prevent any firm confirmation.

        \subsubsection{Fraction of quenched galaxies}\label{sec:f_quenched}

\begin{figure}
            \centering
            \includegraphics[width=\columnwidth]{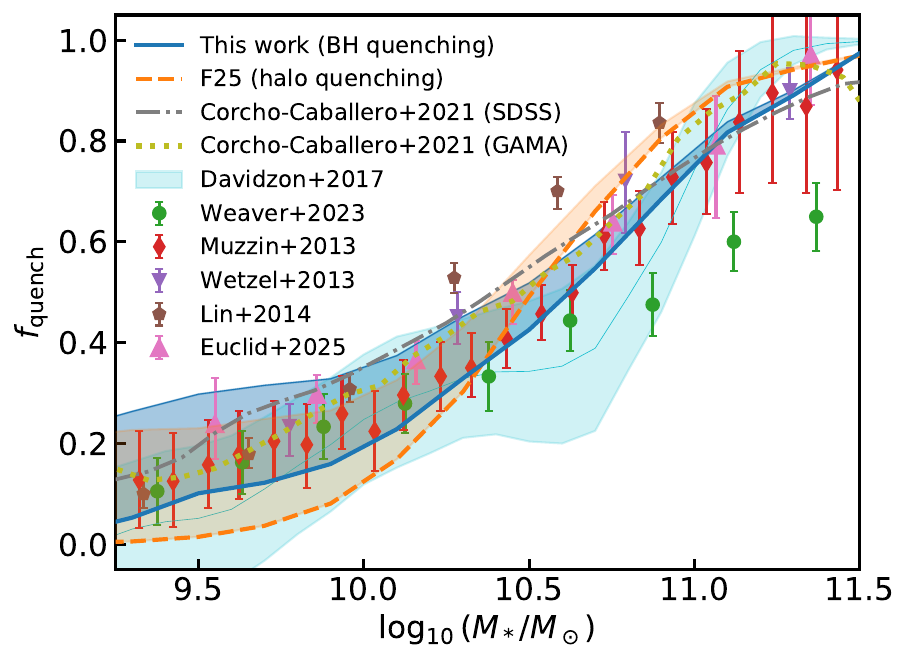}
            \includegraphics[width=\columnwidth]{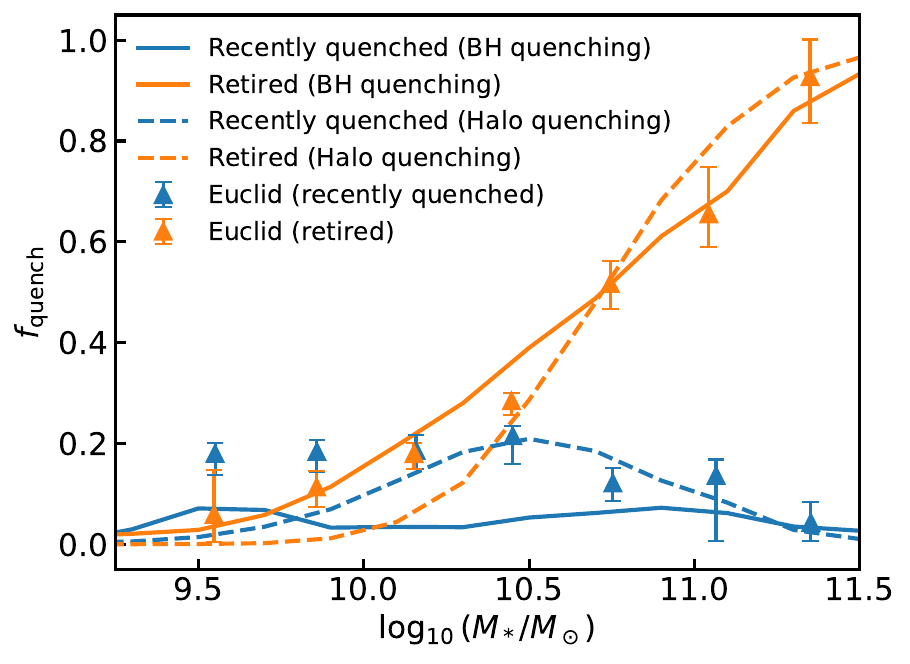}
            \caption{Upper panel: Fraction of quenched galaxies as a function of stellar mass at redshift $z=0$. Blue solid and orange dashed lines show the black hole and halo quenching scenarios, respectively. The blue and orange thin lines and shaded areas denote the upper bound and uncertainty in the extreme assumption of all satellites being quenched. We compare \decode's predictions to the observed quenched fractions from SDSS/GAMA (grey dash-dotted and yellow dotted lines; \citealt{corcho_caballero_2021}), COSMOS2015 (cyan solid lines and shaded areas; \citealt{davidzon_2017}), COSMOS2020 (green dots with error bars; \citealt{weaver_2023}), \citet{muzzin_2013} (red rhombuses with error bars), \citet{wetzel_2013} (purple triangles with error bars), \citet{lin_2014} (brown pentagons with error bars) and Euclid (pink triangles with error bars; \citealt{corcho_caballero_2025}). Lower panel: Same as upper panel, but divided into recently quenched (blue lines and triangles) and retired (orange lines and triangles) galaxies. We compare our results to the fraction of quenched and retired galaxies from the Euclid Q1 data release (\citealt{corcho_caballero_2025}).}
            \label{fg:fquenched}
        \end{figure}

        In this section, we present the fractions of quenched galaxies as predicted by the black hole quenching model (blue lines) and, as a term of comparison, also the halo quenching model introduced by \citetalias{fu_2025} (orange lines). The upper Figure \ref{fg:fquenched} reports the fraction of quenched galaxies as a function of stellar mass predicted by our model at $z=0$ compared to those observationally inferred from SDSS/GAMA (grey dash-dotted and yellow dotted lines; \citealt{corcho_caballero_2021}), COSMOS2015 (cyan solid lines and shaded areas; \citealt{davidzon_2017}), COSMOS2020 (green dots with error bars; \citealt{weaver_2023}), \citet{muzzin_2013} (red rhombuses with error bars), \citet{wetzel_2013} (purple triangles with error bars), \citet{lin_2014} (brown pentagons with error bars) and Euclid (pink triangles with error bars; \citealt{corcho_caballero_2025}). We stress that our quenching recipe applies only to the evolution of central galaxies for which we can predict the number of quenched galaxies in each stellar mass bin (blue and orange thick lines). However, such a prediction constitutes a lower limit, since it does not contain the population of quenched satellite galaxies. We thus also show the extreme scenario in which all the surviving (unmerged) galaxy satellites at $z=0$ are quenched (blue and orange thin lines) to provide a robust upper limit to the total fraction of quenched galaxies predicted by our models. The overall trend of the observed quenched fractions is well reproduced by both the halo and black hole quenching models, which continue to remain quite degenerate at this level of the analysis.

        To further test the predictions of our quenching models, we also compare with the latest findings from the Euclid Collaboration on quenched galaxies. The lower panel of Figure \ref{fg:fquenched} shows the fraction of quenched galaxies as a function of stellar mass at $z=0$ segregated into recently quenched galaxies and retired galaxies, as derived by \citet{corcho_caballero_2025}, with sSFRs averaged over 100 Myr (${\rm sSFR}_8$) and 1 Gyr (${\rm sSFR}_9$), respectively. We report below the definitions of recently quenched galaxies (QGs) and retired galaxies (RGs) from Equations (2) and (3) in \citet{corcho_caballero_2025} for compactness:
        \begin{equation}
            {\rm QGs} \left\{
            \begin{aligned}
            & \log_{10} \, {\rm sSFR}_9 > -11 \; , \\
            & \log_{10} \, {\rm sSFR}_8 < -11 \; , \\
            & \log_{10} \, {\rm sSFR}_8 < \log_{10} \, {\rm sSFR}_9 - 1 \; ,
            \end{aligned}
            \right .
        \end{equation}
        \begin{equation}
            {\rm RGs} : \left\{
            \begin{aligned}
                & \log_{10} \, {\rm sSFR}_9 < -11 \; , \\
                & \log_{10} \, {\rm sSFR}_8 < -11.5 \; .
            \end{aligned}
            \right.
        \end{equation}
        We find that whilst both quenching models tend to broadly reproduce the overall fraction of retired galaxies, in line with the upper panel of Figure \ref{fg:fquenched}, there is a tendency for black hole quenching scenario to generate a lower number of recently quenched galaxies at $M_\star \lesssim 10^{11}\, M_\odot$ than actually observed, with the halo quenching model performing better. However, given the systematic uncertainties in the predicted fractions of quenched satellites, both models could still end up aligning with the data.

        \subsubsection{The black hole mass-stellar mass relation}\label{sec:Mbh_Mstar}

        \begin{figure}
            \centering
            \includegraphics[width=0.978\columnwidth]{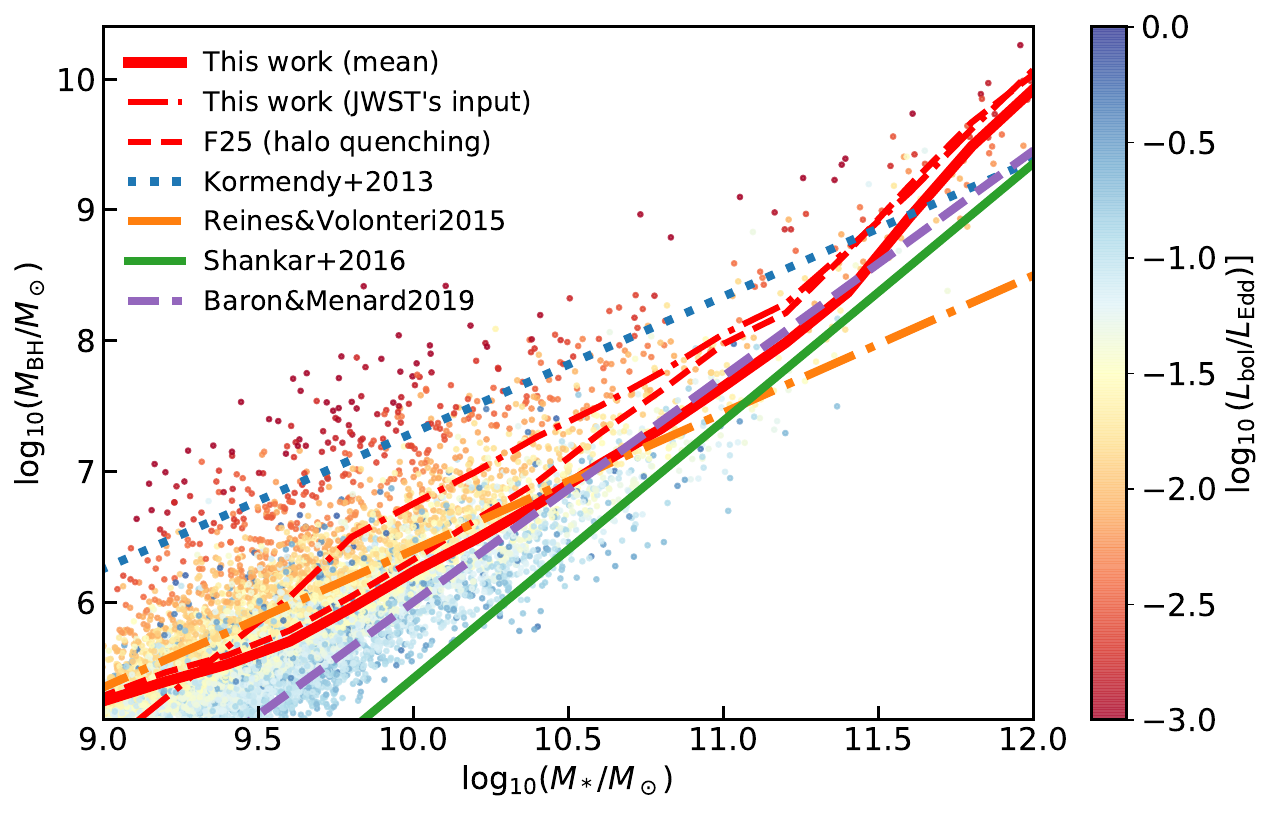}
            \hspace*{0.045cm}
            \includegraphics[width=\columnwidth]{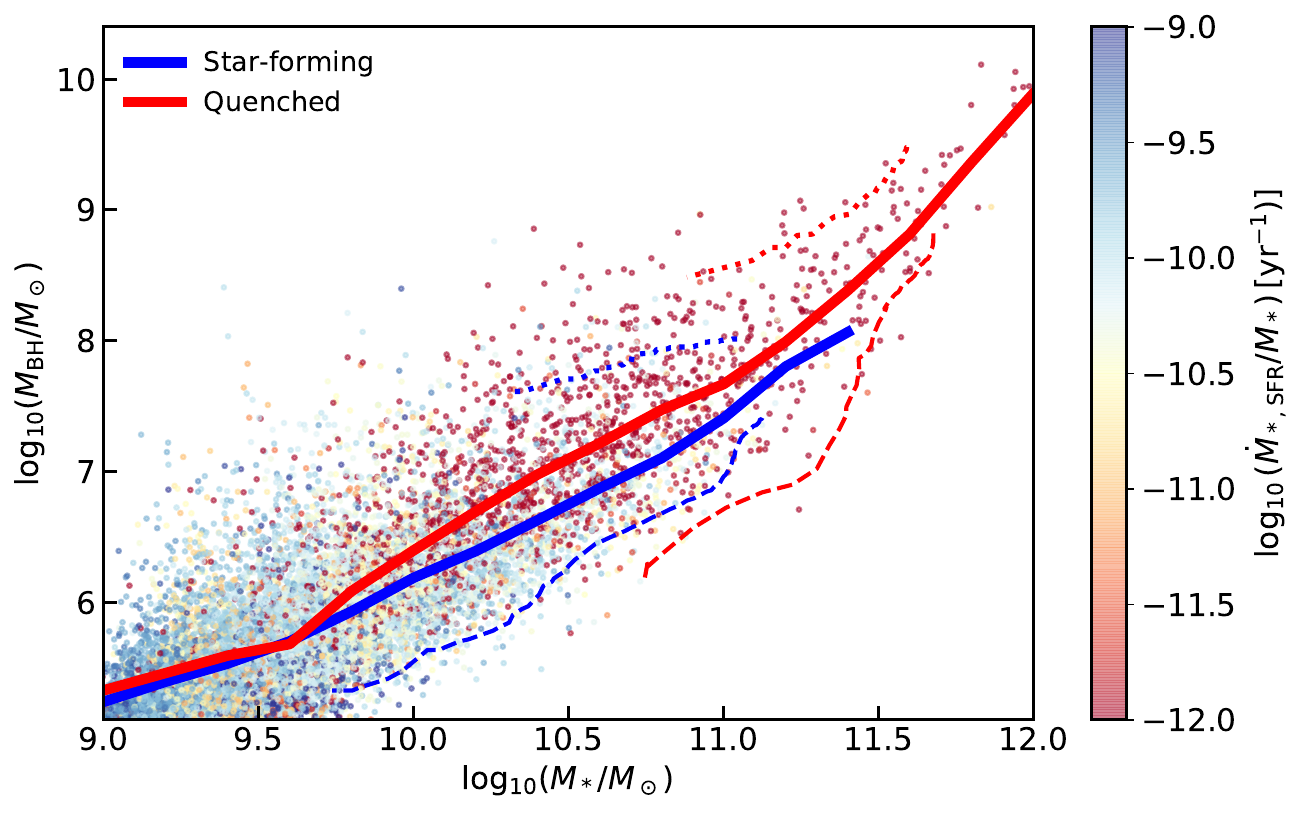}
            \caption{Upper panel: Local black hole mass-stellar mass relation predicted by \decode colour-coded with the $L_{\rm bol} / L_{\rm Edd}$ ratio. The red solid and dashed lines depict the mean relations from \decode for the black hole quenching (this work) and halo quenching (\citetalias{fu_2025}) scenarios, respectively. The red dash-dotted line show the case of adopting the JWST \citet{pacucci_2023} $M_{\rm BH}-M_\star$ seeding. \decode's results are compared to the observationally measured relations from \citet[][blue dotted line]{kormendy_ho_2013}, \citet[][orange dash-dotted line]{reines_volonteri_2015}, \citet[][green solid line and shaded area]{shankar_2016} and \citet[][purple dashed line]{baron_menard_2019}. Lower panel: Same as upper panel, but colour-coded with the galaxy specific star formation rate. The blue and red dashed lines show the average mass evolution of black holes selected in star-forming galaxies with stellar mass $10^{11}\, M_\odot < M_\star < 10^{11.2} \, M_\odot$ and quenched galaxies with stellar mass $M_\star > 10^{11.5} \, M_\odot$, respectively, approaching the mean relation from below. The blue and red dotted lines show the same quantity but for black holes approaching the mean relation from above.}
            \label{fg:Mbh_Mstar_local}
        \end{figure}

        In this section, we study \decode's predictions for the black hole-galaxy scaling relations. Figure \ref{fg:Mbh_Mstar_local} shows the black hole mass-stellar mass ($M_{\rm BH}-M_\star$) relation in the local Universe predicted by \decode as a function of the AGN bolometric luminosity-to-Eddington luminosity ratio (upper panel) and galaxy sSFR (lower panel). In the upper panel, we compare \decode's results to several observationally inferred relations in the literature (\citealt{kormendy_ho_2013, reines_volonteri_2015, shankar_2016, baron_menard_2019}). In the lower panel, we show our $M_{\rm BH}-M_\star$ relation divided for star-forming and quenched galaxies.

        First of all, we note that the trends shown in Figure \ref{fg:Mbh_Mstar_local} also emerge from the halo quenching model of \citetalias{fu_2025}. Indeed, both black hole and halo quenching models produce similar stellar mass functions, stellar mass-halo mass relations, fractions of quenched galaxies, and even star formation histories (as discussed in the previous sections). Since \decode's black hole accretion histories are built following the star formation histories of their host galaxies, in turn, also the output black hole accretion histories and implied $M_{\rm BH} - M_\star$ scaling relations will be similar, as shown in the upper panel of Figure \ref{fg:Mbh_Mstar_local}.

        We found that, on the assumption of a standard $\varepsilon \sim 0.1$, our predicted local $M_{\rm BH}-M_\star$ relation is aligned with those from \citet{reines_volonteri_2015} and \citet{baron_menard_2019}, who measured the $M_{\rm BH}-M_\star$ relation for active galaxies selected from the Sloan Digital Sky Survey. We also observe that in our models at fixed stellar mass, more massive SMBHs favour quenched host galaxies that live above the mean relation, as shown in the lower panel, with quenched galaxies living on a mean $M_{\rm BH}-M_\star$ relation of $0.3-0.4$ dex higher than that of the star-forming galaxies on average (at least above $M_\star \gtrsim 6 \times 10^9 \, M_\odot$), consistent with what retrieved in simulations (see e.g. \citealt{terrazas_2016, terrazas_2020}). A more marked correlation is present for the $L_{\rm bol}/L_{\rm Edd}$ ratio, with lower values of luminosity also lying above the mean relation, similar to the distribution of the sSFR, reflecting the fact that quiescent mode of the black hole accretion follows the drop of the galaxy SFR. The predicted strong segregation in Eddington ratio in the $M_{\rm BH}-M_\star$ plane may contribute to the observed discrepancies in the normalisations of the $M_{\rm BH}-M_\star$ relations characterising active and normal galaxies (e.g. \citealt{reines_volonteri_2015, shankar_2019}).

        The assumed radiative efficiency modulates the normalisation of the $M_{\rm BH}-M_\star$ relation, resulting in a lower relation for any value of $\varepsilon\geq0.1$. We checked that for values up to $\varepsilon=0.15-0.2$ our $M_{\rm BH}-M_\star$ relation would be $\sim 0.5-0.8$ dex lower in normalisation with respect to the one calibrated on the local dynamical SMBHs, which could instead be reproduced assuming $\varepsilon \sim 0.02$ (see also \citealt{yang_2019}, \citealt{shankar_2020} and \citealt{zou_2024}).

        In the lower panel of Figure \ref{fg:Mbh_Mstar_local}, we also show the average evolution of black holes selected in star-forming galaxies with stellar mass $10^{11}\, M_\odot < M_\star < 10^{11.2} \, M_\odot$ (blue dashed and dotted lines) and quenched galaxies with stellar mass $M_\star > 10^{11.5} \, M_\odot$ (red dashed and dotted lines), approaching the local $M_{\rm BH}-M_\star$ relation from above and below. We find that SMBHs hosted in star-forming or quiescent galaxies share similar evolutionary tracks. In general, the way black holes reach the mean $M_{\rm BH}-M_\star$ relation is more sensitive to the initial condition. If the SMBHs start at $z=6$ from a relatively low mass baseline, set by either the \citet{reines_volonteri_2015} relation or the $10^5 \, M_\odot$-seeding model, then they will preferentially approach the $M_{\rm BH}-M_\star$ relation from below (dashed lines). On the other hand, if SMBHs are initialised at $z=6$ on the \citet{pacucci_2023} relation, many (but not all) of them will approach the relation also from above.

        \begin{figure}
            \centering
            \includegraphics[width=\columnwidth]{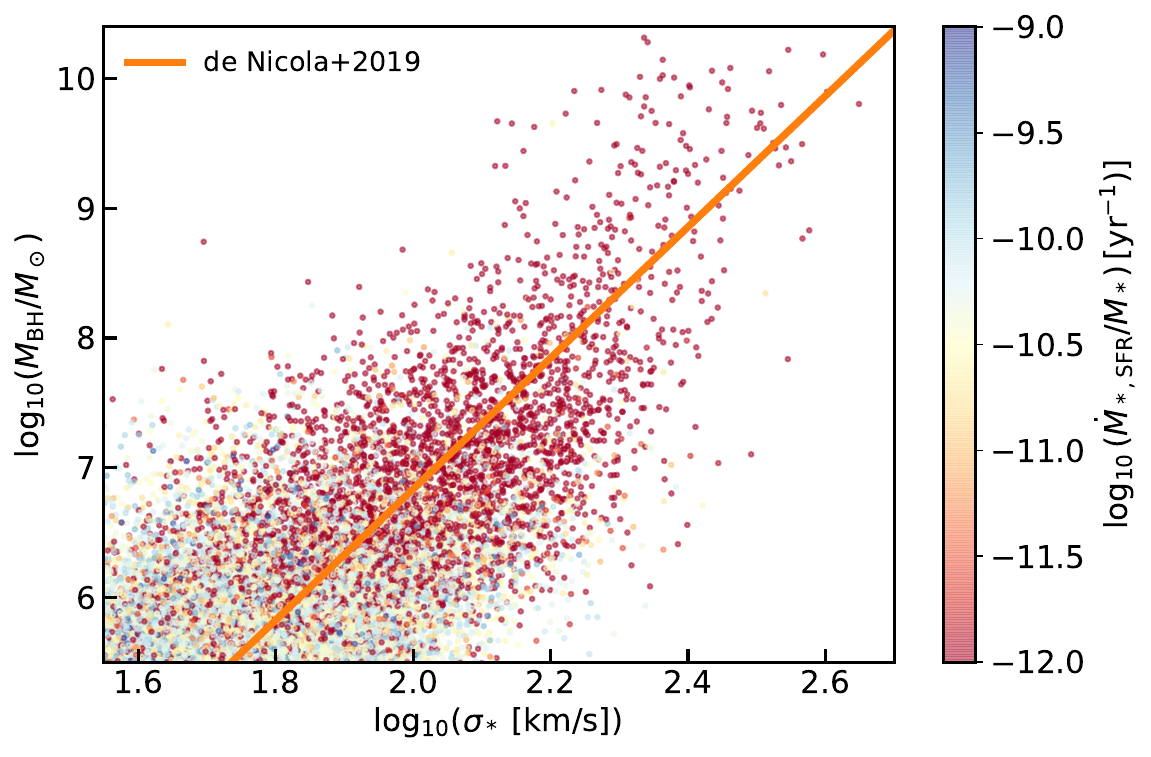}
            \caption{Black hole mass-stellar velocity dispersion relation at $z=0$ as a function of specific star formation rate, compared to the observed relation from \citet[][orange solid line]{de_nicola_2019}.}
            \label{fg:Mbh_sigma}
        \end{figure}

        Figure \ref{fg:Mbh_sigma} reports our predicted $M_{\rm BH} - \sigma_\star$ relation compared to the one by \citet{de_nicola_2019}, which is based on the same aperture and which we adopt as a reference to trigger the quenching of host galaxies. We find that all our galaxies nicely cluster around this relation and have an average consistent with the observed one for standard values of $\varepsilon \sim 0.1$, at variance with what retrieved in Fig \ref{fg:Mbh_Mstar_local} for the $M_{\rm BH} - M_\star$, once more pointing to the universality and robustness of the $M_{\rm BH} - \sigma_\star$ relation (e.g. \citealt{shankar_2025} and references therein). We also find absent redshift-evolution in the $M_{\rm BH} - \sigma_\star$ relation both in normalisation and slope.

        It is also relevant to highlight that the choice of initial condition does not have a major impact on the subsequent timeline for quenching. Regardless of the seeding of the black holes, most of the galaxies reach the local $M_{\rm BH} - \sigma_\star$ from below and quench their star-formation afterwards, even when adopting the JWST high $M_{\rm BH}-M_\star$ relation from \citet{pacucci_2023} or \citet{li_2025}. In fact, in the latter case, when starting our simulation (at about $z\sim 6$), the black holes are initialised with a mass above the mean $M_{\rm BH} - \sigma_\star$ due to the higher normalisation of the high-$z$ $M_{\rm BH}-M_\star$ relation. However, after having set the initial condition, we do not immediately apply the $M_{\rm BH} - \sigma_\star$ quenching condition, as their position on the relation may not correspond to their real evolutionary stage at that redshift. In a few Myr time (by $z\gtrsim 4-5$), all galaxies grew rapidly via star formation moving them below the mean $M_{\rm BH} - \sigma_\star$ and from then they followed the same typical evolutionary path as in our reference model, yielding similar results at low redshifts.

        \begin{figure}
            \centering
            \includegraphics[width=\columnwidth]{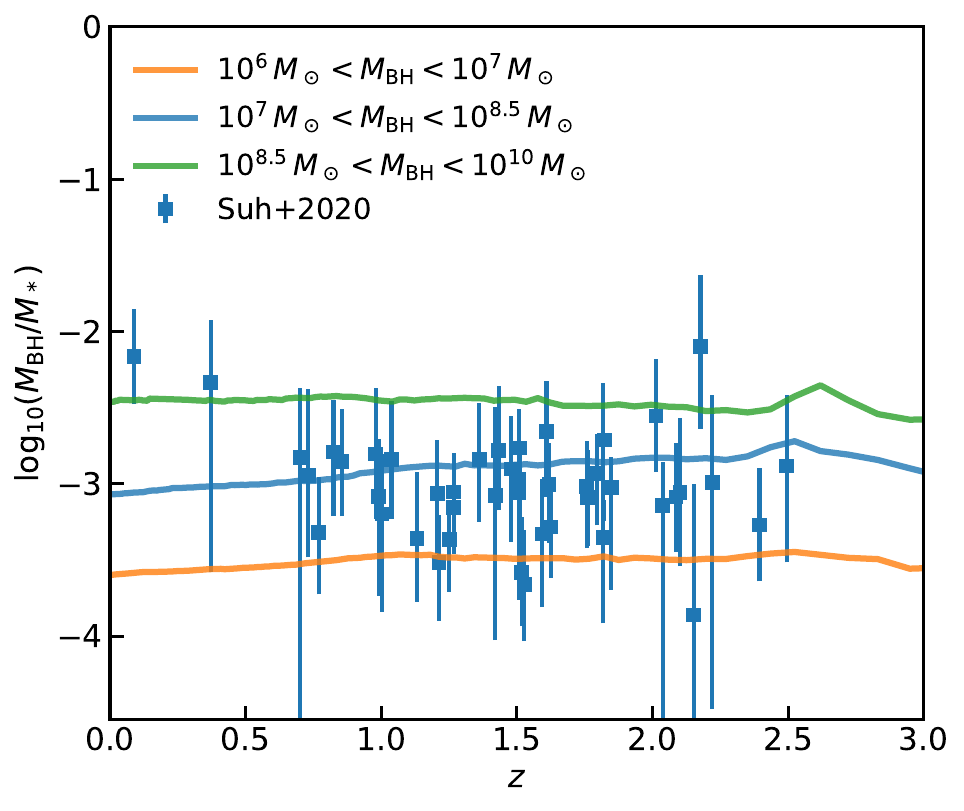}
            \caption{Black hole mass-stellar mass ratio as a function of redshift predicted by \decode (coloured solid lines) for black holes selected in different mass bins. Data points and error bars show the observations from the Chandra-COSMOS Legacy Survey (\citealt{suh_2020}) for black holes with mass $10^{7} \, M_\odot < M_{\rm BH} < 10^{8.5} \, M_\odot$.}
            \label{fg:Mbh_Mstar_ratio}
        \end{figure}

        We also find nearly no evolution in redshift for the $M_{\rm BH}-M_\star$ relation, in line with what shown by other models (e.g. \citealt{shankar_2020}) and observations. Figure \ref{fg:Mbh_Mstar_ratio} shows the redshift-evolution of the ratio between black hole mass and host galaxy stellar mass, for different black hole mass bins ($10^{6} \, M_\odot < M_{\rm BH} < 10^{7} \, M_\odot$, $10^{7} \, M_\odot < M_{\rm BH} < 10^{8.5} \, M_\odot$ and $10^{8.5} \, M_\odot < M_{\rm BH} < 10^{10} \, M_\odot$). We compare our predictions with the evolution inferred in \citet{suh_2020} from the Chandra-COSMOS Legacy Survey AGN catalogue. Similar findings are reported also by other observations and models, although we do not show them in the comparison due to the different stellar mass samples (e.g. \citealt{merloni_2010, sun_2015, cloonan_2024, dattathri_2025, sun_2025a, sun_2025b}). Overall, we found no significant evolution for the $M_{\rm BH}-M_\star$ relation, with the value of the $M_{\rm BH}/M_\star$ ratio confined between $\sim 0.01 \%$ and $1 \%$, and a variation of $\lesssim 5 \%$ up to redshift $z<3$, relatively consistent with what found in \citet{suh_2020} and other works.

\section{Discussion}\label{sec:discuss}

The aim of this work is to test the viability of the feedback from SMBHs in driving the fractions of quenched galaxies as well as other related observables, such as the $M_{\rm BH} - M_\star$ and the $\dot M_{\rm BH} - M_\star$ relations. To this purpose, we adopted a transparent and data-driven approach to build the population of central galaxies and SMBHs across cosmic time based on empirical SFR-HAR relations and Eddington ratio distributions to assign SFRs and BH accretion rates, respectively, to galaxies along their evolutionary tracks, thus progressively building their stellar and BH masses. Here we discuss some of the main relevant aspects of our approach, including some of their limitations and implications.

First of all, we have presented the SFR function at each redshift using both UV and IR data. We note that the inclusion of the IR data could potentially lead to overestimating the SFRs due to the diffuse dust coming from older stars. In this respect, the dust emission from UV light would represent a more accurate tracer of the SFR (see also discussion in \citealt{mancuso_2016}). Correcting the SFRs to account for the diffuse dust emission lies beyond the purpose of this work and several groups have also discussed that such an effect is negligible in galaxies with $\log_{10} \, \dot M_{\rm \star, SFR} \gtrsim 1.5-2$ (e.g. \citealt{clemens_2013, rowlands_2014, swinbank_2014, da_cunha_2015, da_cunha_2021, maresca_2022, sun_2022, sharma_2024}). Anyhow, we tested that if we neglect the IR data and consider only the UV as a tracer of the SFR in galaxies, our abundance matching yields a SFR-HAR relation $1-2$ dex lower in normalisation than in our reference model which would drastically fall short in reproducing the local population of galaxies even without any quenching.

Our data-driven models naturally predict that the SFR and BHAR tracks peak at around the same epochs, suggesting a close link between these two processes. In particular, in our model the galaxy SFR drops with a timescale of roughly $\sim 1$ Gyr, whereas the black hole accretion declines with a timescale larger by a factor of $2-3$. This finding aligns with the empirical findings by \citet{lapi_2014}, who investigated the co-evolution of SMBH and massive galaxies using FIR selected galaxies and X-ray/optically selected AGNs. In particular, \citet{lapi_2014} found evidence for massive galaxies halting their star formation abruptly over $\sim 1$ Gyr possibly due to quasar feedback, with the central SMBHs gradually reducing its mass accretion rate on a slightly longer timescale.

We have also found a delay of $0.5 - 1$ Gyr between the peak of the SFR and the peak of the BHAR. This finding is supported by the results of hydrodynamical simulations such as MACER (\citealt{yuan_2018, zhu_2023, zhang_2025}). MACER is a suit of zoomed-in two-dimensional simulation designed to study isolated galaxies by resolving the Bondi radius and incorporating state-of-the-art AGN physics recipes. The MACER simulation have also found an average time lap of $\sim 0.7$ Gyr between the SFR and BHAR peaks, due to the infalling gas from the circumgalactic medium to the galactic disc enhancing star formation and then propagating towards the galactic centre triggering the black hole accretion. Afterwards, the energy released by the black hole heats the cold gas quenching the star formation of the host galaxy.

Our results are based on the assumption of a redshift-independent $M_{\rm BH}-\sigma_\star$ relation as input threshold for the black hole quenching. While several works have reported nearly absent time evolution for the $M_{\rm BH}-\sigma_\star$ relation (e.g. \citealt{shankar_2009}), other works found non-negligible evolution for the latter. For example, \citet{salviander_2013} showed that, for massive black holes of $M_{\rm BH} \gtrsim 10^9 \, M_\odot$, the stellar velocity dispersion decreases with increasing redshift, even though such evolution does not impact our results since the presence of such massive black holes at high redshifts is extremely rare. However, this apparent redshift-evolution from the local relation is due to observational biases and uncertainties (see discussion in Sect. 5 of \citealt{salviander_2013}) and, therefore, we assumed a purely non-evolving $M_{\rm BH}-\sigma_\star$ scenario in our modelling.

All in all, we find that the black hole quenching and halo quenching models are quite degenerate between each other. In both models we grew the SMBHs via the same empirical $P(\lambda)$ distributions but quenched the galaxies in one instance when the SMBH reaches the $M_{\rm BH}-\sigma_\star$ ridge and in the other when the host halo mass surpasses the golden mass threshold. The predicted quenched fractions appear similar in both models, as well as the implied SMBH scaling relations and, as expected, also the input accretion rates and SFRs are similar in both models (depending on the HAR). Additionally, we find that the $\dot{M}_{\rm BH}-M_\star$ relation is strongly consistent between the two models and with the observed relations (Figure \ref{fg:BHAR_Mstar}), further underlining the degeneracy between the two quenching models. Our results suggest that cumulative statistical indicators such as SMBH scaling relations and quenched fractions are not ideal to discern the physical processes behind quenching, at least below $z<4$. Attempting to extend our black hole quenching predictions to $z>4$ is challenging as comprehensive Eddington ratio distributions are still sparse at those early epochs, and they are further complicated by the allegedly significant, but still unclear, contribution from little red dots (e.g. \citealt{akins_2024, kokorev_2024, matthee_2024, greene_2025}). More insightful constraints to discern between SMBH and halo quenching may come from AGN clustering (as mentioned in Sect. \ref{sec:BHfeedback_SFH}), especially if measured as a function of different galactic and AGN properties (e.g. \citealt{allevato_2021}), and from direct analysis of the measured scaling relations between black holes and their hosts. Machine learning regression algorithms indeed favour black hole mass/stellar velocity dispersion as a key predictor of quiescence in galaxies (e.g. \citealt{bluck_2020}), and pairwise residuals also point to stellar velocity dispersion/host potential well as intimately linked to SMBH mass (e.g. \citealt{shankar_2025}), in line with what expected from AGN feedback, further supported by increasing evidence of  multi-scale AGN outflows  (e.g. \citealt{fiore_2017, mussimenta_2023}).

Finally, we note that the \citet{bongiorno_2016} $P(\lambda)$ distributions depend on stellar mass. To disentangle possible serendipities in the black hole accretion-star formation connection introduced implicitly by such dependence, we also tested our model by taking as input the stellar mass-independent $P(\lambda)$ from X-ray detected sources in the Chandra and XMM-Newton surveys (\citealt{georgakakis_2017}). We found, when employing mass-independent accretion rate distributions, very similar results both quantitatively and qualitatively to those obtained by using the mass-dependent distributions, supporting the robustness of our results. \citet{georgakakis_2017} also showed that a SFR- and mass-dependent distribution, such as the one put forward by \citet{aird_2018}, is on average consistent with their mass-independent one. We also note that the X-ray selected AGN sample may be slightly affected by incompleteness due to the obscuration effects, corresponding to column densities of $N_{H} \gtrsim 10^{22}-10^{23} \, {\rm cm}^{-2}$. To account for the effect of the obscured AGNs, explicit dependence of the $P(\lambda)$ on the column density should be included in the model at each redshift (e.g. \citealt{laloux_2024}), which we leave as a future work.

\section{Conclusions}\label{sec:conclu}

In this paper, we have presented our cosmological semi-empirical model \decode to study the co-evolution of SMBHs with their host galaxies and probe the origin of quenching via SMBH feedback and the shaping of SMBH scaling relations through cosmic time. To this purpose, we grew SMBHs following the star formation histories of their host galaxies via empirically measured Eddington ratio distributions $P(z, L_{\rm bol}/M_\star)$ used as inputs to assign accretion rates to each host galaxy of given stellar mass $M_\star$ at any redshift $z$. We then computed SMBH masses by integrating accretion rates ($\dot{M}_{\rm BH} \propto L_{\rm bol}/c^2$) along the host galaxy evolutionary tracks. The SFRs of the host galaxies were assigned via a monotonic relation between SFR and host HAR, as suggested by state-of-the-art hydrodynamical simulations (see Sect. 3 of \citetalias{fu_2025}). When the SMBH mass surpasses the $M_{\rm BH} - \sigma_\star$ threshold we then assumed that the host galaxy permanently quenches. We used this data-driven phenomenological model to make predictions on a variety of independent observables.

Our main results can be summarised as follows:
\begin{itemize}
    \item The simple assumption of imposing the $M_{\rm BH} - \sigma_\star$ condition is sufficient to generate a fraction of quenched galaxies consistent with observations (see Figure \ref{fg:fquenched}), with possibly a slight tendency to under-predict the fractions of recently quenched galaxies as recently measured by Euclid (Figure \ref{fg:fquenched}).
    \item SMBHs acquired most of their mass between redshift $z\gtrsim1-2$ accreting at an average $\lambda_{\rm Edd} \sim 0.3$, which is consistent with what was inferred by \citet{kollmeier_2006}, among others (Figure \ref{fg:Eratio_evo}). Mergers have a relatively small contribution to the final mass budget of a few percent (Figure \ref{fg:mbh_acc_tracks}). The resulting evolutionary tracks below $z\sim2$ are nearly independent of the initial condition set at $z>5$.
    \item We find that most SMBHs cross the $M_{\rm BH} - \sigma_\star$ relation around the same time as when the host haloes reach the golden mass of $M_{\rm h} \sim 10^{12}\, M_\odot$, rendering the black hole and halo quenching processes quite degenerate in most of our results, though the halo quenching scenario tends to induce more massive host dark matter haloes for galaxies and SMBHs of similar mass (Figures \ref{fg:ex_mbh_growth}, \ref{fg:mh_acc_sSFR_BHAR} and \ref{fg:mhalo_dist}).
    \item The predicted $M_{\rm BH} - M_\star$ relation is relatively constant with time and below the one calibrated for local dynamical inactive SMBHs if $\varepsilon \geq 0.1$. The quenched galaxies with ${\rm sSFR} <10^{-11} \, {\rm yr}^{-1}$ and low values of bolometric luminosity live typically above the mean relation, and vice versa for the star-forming population (Figures \ref{fg:Mbh_Mstar_local} and \ref{fg:Mbh_Mstar_ratio}).
    \item The $M_{\rm BH} - \sigma_\star$ relation predicted by our accretion model appears more in line with observational estimates for standard values of the radiative efficiency of $\sim 10\%$ (Figure \ref{fg:Mbh_sigma}).
\end{itemize}

In closing, using an observed accretion rate distribution $P(z, L_{\rm bol}/M_\star)$ as input is a powerful way of modelling black hole growth in a semi-empirical fashion following the evolution of their hosts. This method can accurately and rapidly generate black hole mass evolutionary tracks, star formation histories, and key SMBH-galaxy scaling relations. Semi-empirical, data-driven models like \decode represent powerful, complementary tools to calibrate the full assembly history of SMBHs and shed light on the co-evolution with their host galaxies and dark matter haloes, whilst making full use of the newest incoming data from e.g. JWST (\citealt{sabelhaus_2005, gardner_2006}) and Euclid (\citealt{mellier_2025}), as both inputs and outputs of the models.

\begin{acknowledgements}
We thank E. Merlin and M. Bernardi for the useful discussion on the high-$z$ quenched fractions and age of massive nearby galaxies, respectively. We are also grateful to the anonymous referee for their comments that further improved the clarity of the presentation. HF acknowledges support at Fudan University from the Shanghai Super Post-doctoral Excellence Program grant No. 2024008. FY is supported in part by NSFC (grant Nos. 12133008, 12192220, and 12192223). LB acknowledges financial support from the German Excellence Strategy via the Heidelberg Cluster of Excellence (EXC 2181 - 390900948) STRUCTURES. MA is supported at the Argelander Institute f\"ur Astronomie through the Argelander Fellowship. AG acknowledges funding from the Hellenic Foundation for Research and Innovation (HFRI) project "4MOVE-U" grant agreement 2688, which is part of the programme "2nd Call for HFRI Research Projects to support Faculty Members and Researchers". YP acknowledges support from the National Science Foundation of China (NSFC) grant Nos. 12125301, 12192220 and 12192222, and support from the New Cornerstone Science Foundation through the XPLORER PRIZE.
\end{acknowledgements}

\bibliographystyle{aa}
\bibliography{main}


\begin{appendix}

\section{Fit to the star formation rate function}\label{app:fit_sfrf}

\begin{figure}
    \centering
    \includegraphics[width=\columnwidth]{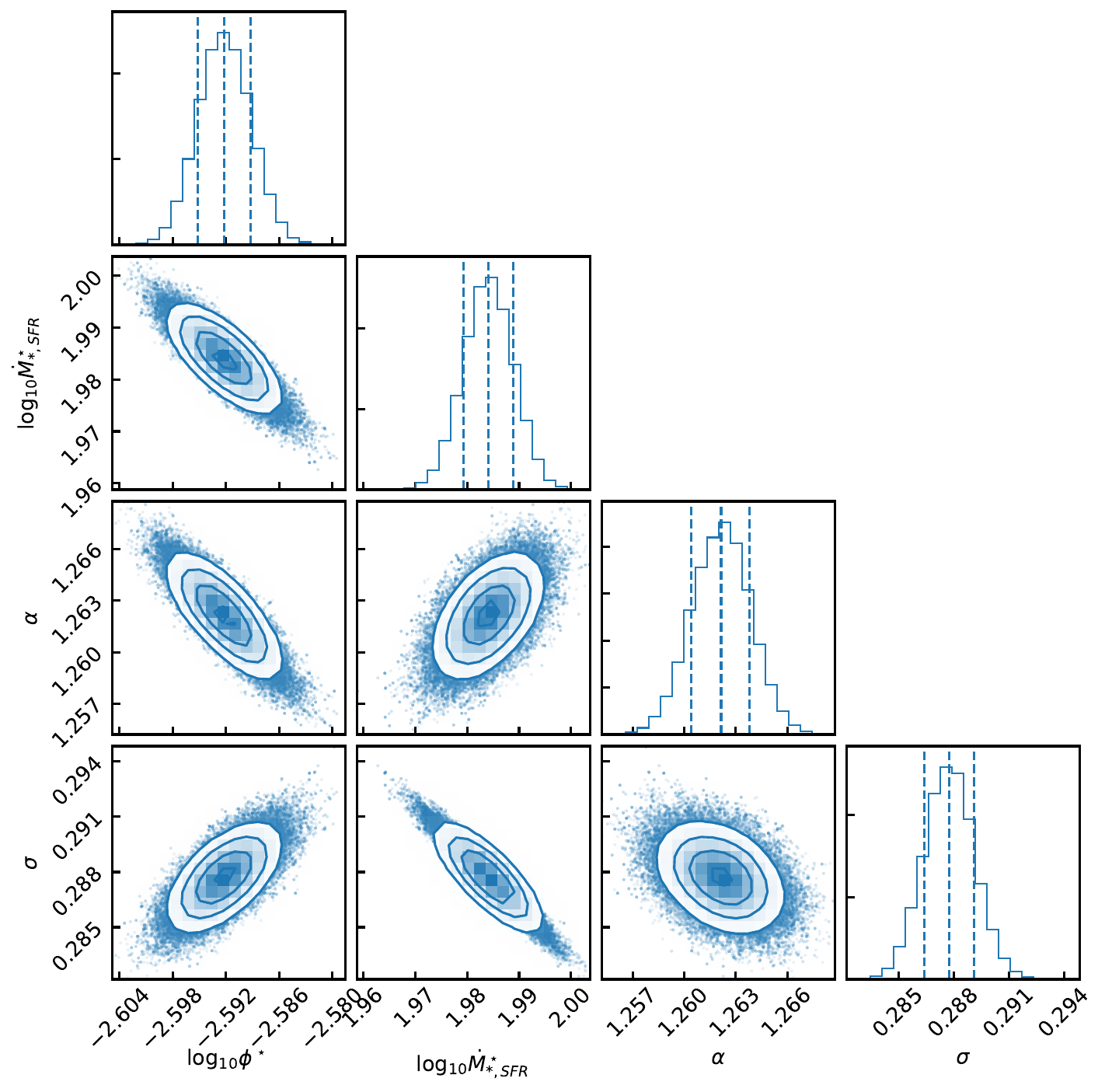}
    \caption{Poster distribution of the fitting parameters to Equation (\ref{eq:saunders}) for the star formation rate function at redshift $z=1$.}
    \label{fg:corner_z1}
\end{figure}

Table \ref{tab:SFR_func_fit_params} shows the best-fitting parameters of the SFR function to Equation (\ref{eq:saunders}) at each redshift. Figures \ref{fg:corner_z1} shows, as example, the posterior distributions of the \citet{saunders_1990} fitting formula's parameters at redshift $z=1$. Similar contour plots are obtained for the fits at other redshifts.

\begin{table}
    \centering
    \begin{tabular}{ccccc}
        \hline
        $z$ & $\log_{10} \phi^\star$ & $\log_{10} \psi_{\rm \star,SFR}$ & $\alpha$ & $\sigma$ \\
        \hline
        $0$ & $-3.576^{+0.018}_{-0.009}$ & $3.943^{+0.028}_{-0.060}$ & $1.302^{+0.001}_{-0.001}$ & $0.001$ \\
        \hline
        $0.5$ & $-2.315^{+0.007}_{-0.007}$ & $0.941^{+0.012}_{-0.012}$ & $1.324^{+0.004}_{-0.004}$ & $0.464^{+0.004}_{-0.004}$ \\
        \hline
        $1$ & $-2.592^{+0.003}_{-0.003}$ & $1.984^{+0.005}_{-0.005}$ & $1.262^{+0.002}_{-0.002}$ & $0.288^{+0.001}_{-0.001}$ \\
        \hline
        $1.5$ & $-3.012^{+0.004}_{-0.004}$ & $2.453^{+0.006}_{-0.006}$ & $1.416^{+0.001}_{-0.001}$ & $0.225^{+0.002}_{-0.002}$ \\
        \hline
        $2$ & $-3.072^{+0.003}_{-0.003}$ & $2.449^{+0.004}_{-0.004}$ & $1.434^{+0.001}_{-0.001}$ & $0.303^{+0.001}_{-0.001}$ \\
        \hline
        $2.5$ & $-3.249^{+0.003}_{-0.003}$ & $2.673^{+0.004}_{-0.004}$ & $1.432^{+0.001}_{-0.001}$ & $0.274^{+0.001}_{-0.001}$ \\
        \hline
        $3$ & $-3.275^{+0.005}_{-0.005}$ & $2.285^{+0.006}_{-0.006}$ & $1.591^{+0.001}_{-0.001}$ & $0.441^{+0.002}_{-0.002}$ \\
        \hline
        $4$ & $-4.709^{+0.018}_{-0.017}$ & $3.049^{+0.019}_{-0.019}$ & $1.829^{+0.002}_{-0.002}$ & $0.268^{+0.006}_{-0.006}$ \\
        \hline
        $5$ & $-4.688^{+0.021}_{-0.021}$ & $2.753^{+0.024}_{-0.024}$ & $1.820^{+0.002}_{-0.002}$ & $0.363^{+0.009}_{-0.009}$ \\
        \hline
        $6$ & $-5.044^{+0.027}_{-0.027}$ & $2.744^{+0.027}_{-0.027}$ & $1.917^{+0.002}_{-0.002}$ & $0.425^{+0.012}_{-0.012}$ \\
        \hline
    \end{tabular}
    \caption{Best-fitting parameters to Equation (\ref{eq:saunders}) for the SFR function at redshifts $z\leq 6$.}
    \label{tab:SFR_func_fit_params}
\end{table}

We then fit the redshift evolution of each parameter of the Saunders formula ($\log_{10} \phi^\star, \log_{10} \psi_{\rm \star,SFR}, \alpha, \sigma$), using a $\chi^2$ likelihood function, finding the following best-fitting analytic forms
\begin{equation}
    \begin{split}
        \log_{10} \phi^\star (z) & = -3.58 + 9.65 \cdot \log_{10}(1+z) \\
        & - 23.56 \cdot \log^2_{10}(1+z) + 11.87 \cdot \log^3_{10}(1+z) \; ,
    \end{split}
\end{equation}

\begin{equation}
    \begin{split}
        \log_{10} \psi_{\rm \star,SFR} (z) & = 3.94 - 20.06 \cdot \log_{10}(1+z) \\
        + 51.63 & \cdot \log^2_{10}(1+z) - 35.37 \cdot \log^3_{10}(1+z) \; ,
    \end{split}
\end{equation}

\begin{equation}
    \begin{split}
        \alpha (z) = 1.30 & - 0.55 \cdot \log_{10}(1+z) \\
        & + 2.11 \cdot \log^2_{10}(1+z) - 0.66 \cdot \log^3_{10}(1+z) \; ,
    \end{split}
\end{equation}

\begin{equation}
    \begin{split}
        \sigma (z) = 0.001 & + 2.75 \cdot \log_{10}(1+z) \\
        & - 6.63 \cdot \log^2_{10}(1+z) + 4.73 \cdot \log^3_{10}(1+z) \; .
    \end{split}
\end{equation}

\section{Galaxy and AGN abundances}\label{app:gal_agn_abun}

We show in this Appendix the galaxy stellar mass function (SMF), fraction of quenched galaxies and AGN X-ray luminosity function.

\begin{figure}
    \centering
    \includegraphics[width=\columnwidth]{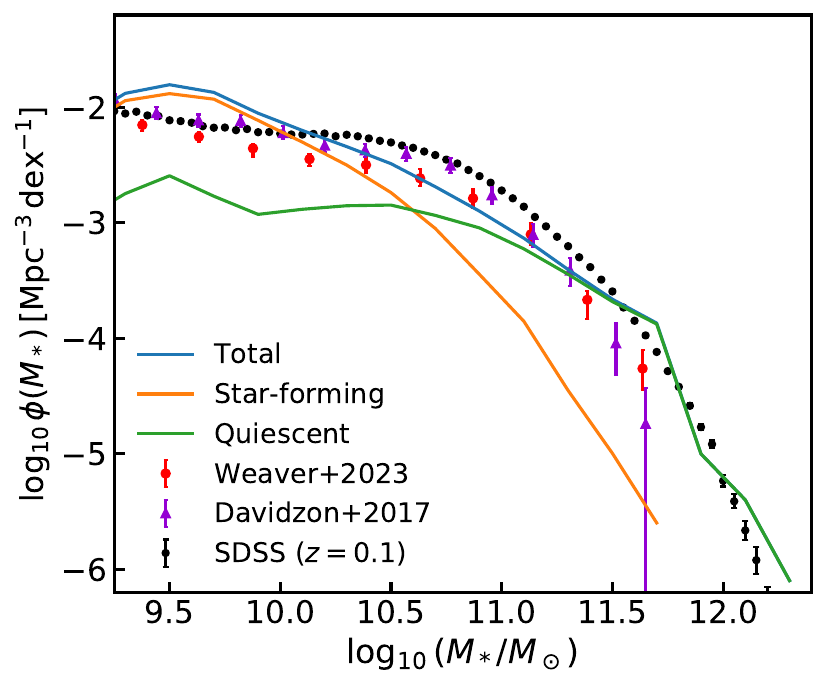}
    \caption{Stellar mass function predicted by \decode at redshift $z=0$. The blue, orange, and green lines show the stellar mass function for all galaxies, star-forming galaxies and quenched galaxies. The grey dots, red circles and purple triangles with error bars show the observational total stellar mass functions from the SDSS at $z=0.1$ (\citealt{bernardi_2017}), COSMOS2020 (\citealt{weaver_2023}) and COSMOS2015 (\citealt{davidzon_2017}) surveys.}
    \label{fg:SMF_fq}
\end{figure}

Figure \ref{fg:SMF_fq} shows the local SMF predicted by \decode, compared to the most recently observationally determined ones (\citealt{bernardi_2017}, \citealt{davidzon_2017} and \citet{weaver_2023}). Our model can fairly well reproduce the observed SMF matching the shape both around the knee and in the massive end. Being able to match the SMF implies the robustness of our star formation histories on top of which our black hole mass assemblies are computed.

\begin{figure}
    \centering
    \includegraphics[width=\columnwidth]{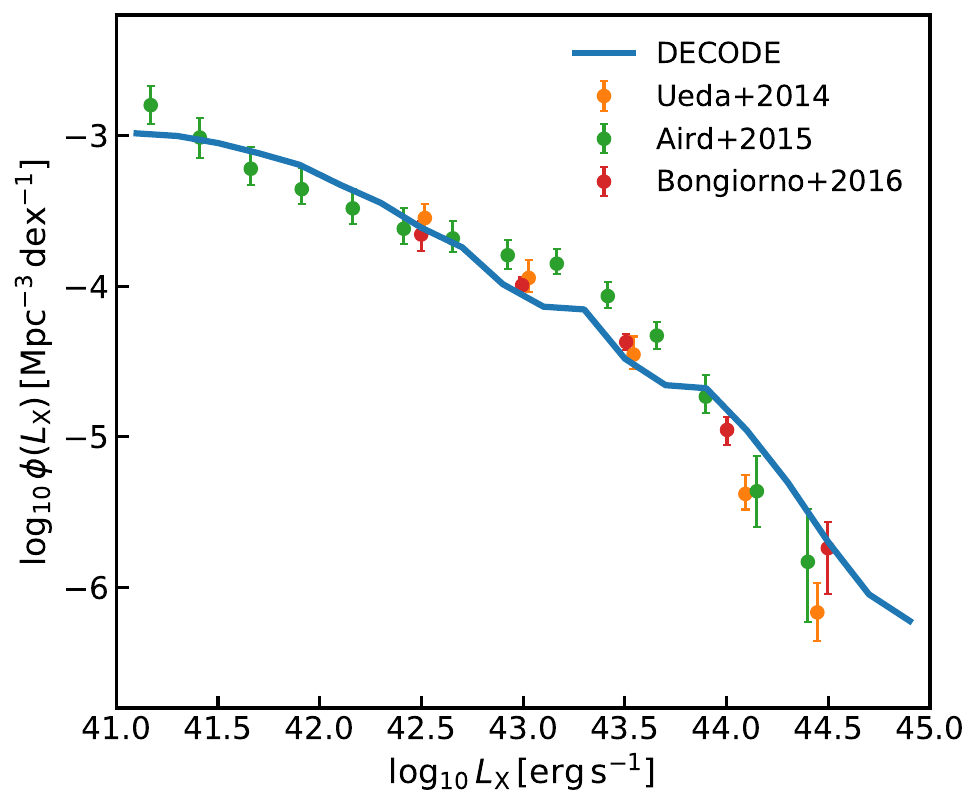}
    \caption{X-ray luminosity function at redshift $z\sim 0.55$ predicted by \decode (blue solid line) compared to those inferred by \citet[][orange dash-dotted line]{ueda_2014}, \citet[][grey solid line and shaded area]{aird_2015} and \citet[][green solid line]{bongiorno_2016}.}
    \label{fg:ANG_LF}
\end{figure}

Finally, Figure \ref{fg:ANG_LF} shows the X-ray LF of AGNs predicted by \decode compared to those of X-ray selected AGNs inferred from different surveys (\citealt{ueda_2014}, \citealt{aird_2015} and \citealt{bongiorno_2016}). We found that the output LF from \decode is in agreement with those observationally inferred, demonstrating the robustness of our predictions on the black hole mass growth and inner self-consistency of the model.

\section{The black hole accretion rate-stellar mass relation}\label{app:Mbhdot_Mstar}

\begin{figure}
    \centering
    \includegraphics[width=\columnwidth]{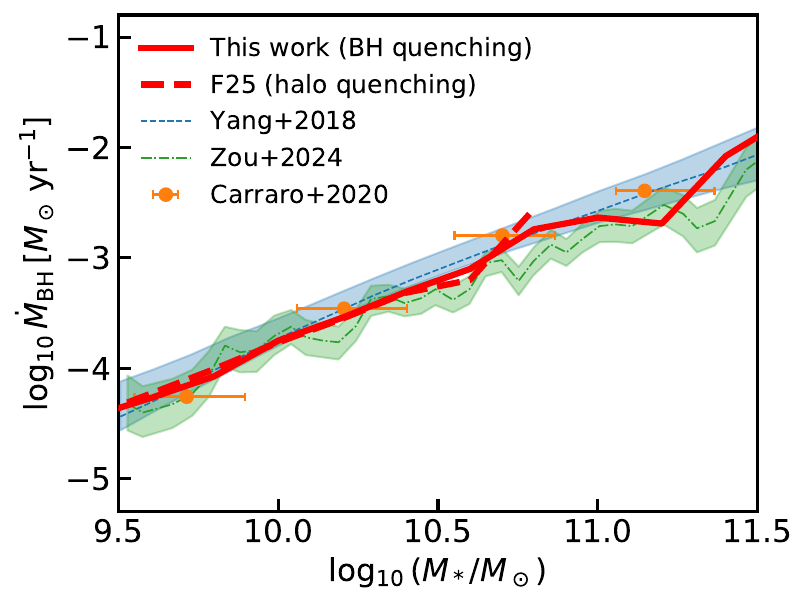}
    \caption{Black hole accretion rate-stellar mass relation at redshift $z\sim 0.375$ for star-forming galaxies. The red solid and dashed lines show the predictions from \decode for the black hole and halo quenching scenarios, respectively. The red dash-dotted line shows the case where we use the JWST \citet{pacucci_2023} relation as input. The blue dashed, green dash-dotted and shaded areas show the relations from \citet{yang_2018} and \citet{zou_2024a}. The dots with error bars show the measurements from \citet{carraro_2020} at redshift $0.1<z<0.65$.}
    \label{fg:BHAR_Mstar}
\end{figure}

Finally, in Figure \ref{fg:BHAR_Mstar} we show the mean BHAR as a function of stellar mass for star-forming galaxies. We compare our prediction to the relation inferred in \citet[][COSMOS, GOODS-N, GOODS-S]{yang_2018}, \citet[][Chandra COSMOS-Legacy survey]{carraro_2020} and \citet[][CANDELS, LSST]{zou_2024a}. Overall, our $\dot{M}_{\rm BH} - M_\star$ relation is consistent with the observational data in normalisation. We find a slope of the relation for the star-forming galaxies of $1.18$, relatively in good agreement with the determinations in the literature. The consistency with the observed $\dot{M}_{\rm BH} - M_\star$ relation yields further support to the validity of our model. We also find that the BHAR as a function stellar mass does not alter in both the halo quenching model (\citetalias{fu_2025}) and black hole quenching model, as a consequence of the fact that the resulting star formation histories, on which we grow our SMBHs, and the relative amounts of star-forming and quenched galaxies are the comparable in the two scenarios.

\end{appendix}

\end{document}